\documentclass[11pt]{article}
\pdfoutput=1

\usepackage{color}
\usepackage{mhchem}
\usepackage{xcolor}
\usepackage{colortbl}
\usepackage{hhline}
\usepackage{cite}
\usepackage{hyperref}
\hypersetup{colorlinks=true,linkcolor=red,anchorcolor=black,citecolor=green}
\usepackage[toc,page]{appendix}
\usepackage{mathtools}
\usepackage{amsfonts}
\usepackage{bbold}
\usepackage{textcomp}
\usepackage[DIV13]{typearea}
\usepackage{amsmath, amsthm, amssymb, mathtools,empheq,latexsym,dsfont}
\usepackage{bbm}
\usepackage{amsmath,amssymb,latexsym,dsfont}
\usepackage{slashed, simplewick}
\usepackage[utf8]{inputenc}
\usepackage{graphicx}
\usepackage{graphicx,placeins}
\usepackage{makeidx}
\usepackage[font=small,labelfont=bf]{caption}
\usepackage{nicefrac}
\usepackage{subfigure}
\usepackage{array, bigdelim,multirow,multicol}
\usepackage{mathrsfs}
\usepackage{adjustbox}
\usepackage{booktabs}
\usepackage{indentfirst}
\usepackage{booktabs}
\usepackage{tabulary}
\usepackage{longtable}
\usepackage{fancybox}
\usepackage{graphicx}
\usepackage{bm}
\usepackage{bbm}
\usepackage{tikz}
\usepackage{diagbox}
\usepackage{enumitem}
\usepackage[integrals]{wasysym}

\graphicspath{{immagini/}}

\definecolor{Gray}{gray}{0.92}

\newcommand{\ignore}[1]{}

\newcommand{\be}{\begin{equation}}
\newcommand{\ee}{\end{equation}}
\newcommand{\bea}{\begin{eqnarray}}
\newcommand{\eea}{\end{eqnarray}}

\definecolor{lightred}{rgb}{1,0.4,0.4}
\definecolor{lightgreen}{rgb}{0.4,1,0.4}
\definecolor{LightCyan}{rgb}{0.88,1,1}

\newcounter{Thm}[section]
\renewcommand{\theThm}{\arabic{section}.\arabic{Thm}}

\newcounter{nodecount}

\tikzstyle{every picture}+=[remember picture,baseline]
\tikzstyle{every node}+=[inner sep=0pt,anchor=base,
text depth=.25ex,outer sep=1.5pt]
\tikzstyle{every path}+=[thick, rounded corners]
\tikzset{
        plabel/.style={inner sep=2pt}
}

\makeatletter
\@addtoreset{equation}{section}
\makeatother

\begin{document}
\title{\begin{center}
{\Large\bf Non-holomorphic modular $A_{5}$ symmetry for lepton masses and mixing}
\end{center}}
\date{}
\author{Cai-Chang Li$^{a,b,c}$\footnote{E-mail: {\tt ccli@nwu.edu.cn}},  \
Jun-Nan Lu$^{d}$\footnote{E-mail: {\tt hitman@mail.ustc.edu.cn}},  \
Gui-Jun Ding$^{d}$\footnote{E-mail: {\tt dinggj@ustc.edu.cn}} \ \\*[20pt]
\centerline{\begin{minipage}{\linewidth}
\begin{center}
$^a${\it\small School of Physics, Northwest University, Xi'an 710127, China}\\[2mm]
$^b${\it\small Shaanxi Key Laboratory for Theoretical Physics Frontiers, Xi'an 710127, China}\\[2mm]
$^c${\it\small NSFC-SPTP Peng Huanwu Center for Fundamental Theory, Xian 710127, China}\\[2mm]
$^d${\it \small Department of Modern Physics and Anhui Center for fundamental sciences in theoretical physics,\\
University of Science and Technology of China, Hefei, Anhui 230026, China}\\[2mm]
\end{center}
\end{minipage}} \\[10mm]}
\maketitle

\thispagestyle{empty}

\centerline{\large\bf Abstract}
\begin{quote}
\indent
We perform a comprehensive bottom-up study of all the simplest lepton models based on non-holomorphic $A_{5}$ modular flavor symmetry, in which neutrinos are assumed to be Majorana particles and their masses are generated by the Weinberg operator or the type I seesaw mechanism. In the case that the generalized CP (gCP) symmetry is not considered, we find that 21 Weinberg operator models and 174 seesaw models can accommodate the experimental data in lepton sector, and all of them depend on six dimensionless free parameters and two overall scales. If gCP symmetry compatible with $A_{5}$ modular symmetry is imposed, one more free parameter would be reduced. Then only 4 of the 21 Weinberg operator models and 100 of the 174 seesaw models agree with the experimental data on lepton masses and mixing parameters. Furthermore, we perform a detailed numerical analysis for two example models for illustration.
\end{quote}
\newpage
\section{Introduction}

The vast majority of free parameters in Standard Model (SM) arise from the flavor sector, including the masses, mixing angles and CP violation phases of quarks and leptons~\cite{ParticleDataGroup:2024cfk}. There are huge hierarchies among the quark masses, for example, the top quark is about five orders of magnitude heavier than the up quark. The quark flavor mixing described by the Cabibbo–Kobayashi–Maskawa (CKM) matrix also exhibits a hierarchical pattern, and all the three quark mixing angles are small. The mass hierarchies in the lepton sector is more dramatic. The neutrinos have a tiny mass and its absolute mass scale is of the order of eV which is approximately nine orders of magnitude smaller than the tau lepton mass. The neutrino oscillation experiments show that the lepton mixing is differ significantly from quark mixing. The solar mixing angle $\theta_{12}$ and the atmospheric mixing angle $\theta_{23}$ are large, while the reactor angle $\theta_{13}$ is much smaller and similar in magnitude to the Cabibbo angle which is the largest quark mixing angle.

The above peculiar flavor structure calls for a dynamical explanation, and lots of effort has been devoted to search a fundamental principle governing the flavor sector of SM. It was found that the large lepton mixing angles can be understood by extending the SM with a non-abelian discrete flavor symmetry~\cite{King:2017guk,Petcov:2017ggy,Xing:2020ijf,Feruglio:2019ybq,Ding:2024ozt}. In the past years, the flavor symmetry based on the modular group opened up a new promising approach to address the flavor structure of SM~\cite{Feruglio:2017spp}. In this paradigm, the three generations of quarks and leptons transform non-trivially under the modular symmetry, and Yukawa couplings are assumed to be modular forms of level $N$ which are holomorphic functions of the complex modulus $\tau$ and non-trivially transform under the action of the modular group. The modular flavor symmetry is formulated in the framework of supersymmetry to ensure the holomorphicity of the Yukawa couplings as functions of $\tau$~\cite{Lauer:1989ax,Ferrara:1989bc,Ferrara:1989qb,Feruglio:2017spp}. The modular symmetry provides a novel origin of the discrete flavor symmetry, and the inhomogeneous (homogeneous) finite modular group $\Gamma_N$ ($\Gamma'_N$) plays the role of flavor symmetry and it is isomorphic to the (double covering of) permutation group for $N\leq5$, e.g., $\Gamma_2\cong S_3$,  $\Gamma_3\cong A_4$,  $\Gamma_4\cong S_4$,  $\Gamma_5\cong A_5$~\cite{deAdelhartToorop:2011re}. The modular flavor symmetry has been intensively studied to explain the quark and lepton flavor structure in the literature, please see the recent reviews of the modular flavor symmetry~\cite{Kobayashi:2023zzc,Ding:2023htn} and references therein. In the minimal modular invariant model to date, the experimental data of all lepton masses and mixing angles can be described in terms of only six free real parameters including $\Re(\tau)$ and $\Im(\tau)$~\cite{Ding:2022nzn,Ding:2023ydy}, and fourteen parameters are invoked to accommodate the masses and mixing parameters of both quark and leptons~\cite{Ding:2023ydy}, in which the complex modulus $\tau$ is a portal so that the flavor observables in quark and lepton sectors are strongly correlated~\cite{Ding:2024pix}. It is notable that modular symmetry models exhibit a universal behavior in the vicinity of fixed points, independently from details of the models~\cite{Feruglio:2022koo,Feruglio:2023mii}. The universality also holds true in multiple moduli models and the same universal scaling behavior as the single moduli case is found~\cite{Ding:2024xhz}. The small departure of the modulus from the fixed points can produce the quark and lepton mass hierarchies~\cite{Okada:2020ukr,Feruglio:2021dte,Novichkov:2021evw,Petcov:2022fjf,Kikuchi:2023cap,Abe:2023ilq,Kikuchi:2023jap,Abe:2023qmr,Petcov:2023vws,deMedeirosVarzielas:2023crv}. Moreover, it was shown that the invariance under the modular group offers an axion-less solution to the strong CP
problem~\cite{Feruglio:2023uof,Petcov:2024vph,Penedo:2024gtb,Feruglio:2024ytl}, and the modulus $\tau$ can play the role of inflaton in the early Universe~\cite{Ding:2024neh,King:2024ssx}.

The modular symmetry is ubiquitous in higher dimensional theories such as superstring theory. For instance, the complex modulus $\tau$ parameterizes the shape of the torus in torus or its orbifold compactification, and the modular symmetry is the geometrical symmetry of the compact space. The top-down approach generally gives rise to both traditional flavor symmetry and modular symmetry. This leads to the idea of eclectic flavor group which is the nontrivial product of the traditional flavor symmetry and modular symmetry~\cite{Baur:2019kwi,Nilles:2020nnc,Nilles:2020kgo,Nilles:2020tdp,Baur:2020jwc,Nilles:2020gvu,Baur:2021mtl,Nilles:2023shk}.
The traditional flavor symmetry severely restricts the K\"ahler potential such that plenty of terms compatible with modular symmetry~\cite{Chen:2019ewa} are suppressed and the flavor universal K\"ahler potential can be produced at leading order. However, one has to reintroduce the notorious flavon fields besides the modulus to break the traditional flavor symmetry. The eclectic flavor group combined the advantages of modular symmetries and traditional flavor symmetries, and some predictive models with eclectic flavor symmetry have been proposed~\cite{Chen:2021prl,Baur:2022hma,Ding:2023ynd,Li:2023dvm,Li:2024pff}.

Recently the non-holomorphic formulation of the modular flavor symmetry has been developed in Ref.~\cite{Qu:2024rns}, and lepton flavor models based on the finite modular group $\Gamma_3\cong A_4$ have been built. The modular invariance is  inherited while the assumption of holomorphicity is superseded by the harmonic condition. As a consequence, the Yukawa couplings are polyharmonic Maa{\ss} forms of level $N$, and supersymmetry is not a requisite anymore. Although the modular weights of non-vanishing modular forms are non-negative\footnote{The modular form of weight zero is a constant which is $\tau$-independent.}, the weights of polyharmonic Maa{\ss} forms can be negative integers. Moreover, there exist non-holomorphic polyharmonic Maa{\ss} forms apart from the known holomorphic modular forms at weight one and weight two. This formalism can be extended to consistently include generalized CP (gCP) symmetry which constrains the phases of the coupling constants, thus the predictive power of the modular invariant models are improved further, as in the supersymmetric modular flavor symmetry~\cite{Novichkov:2019sqv}.

The non-holomorphic modular invariant approach to the flavor problem has been explored at levels $N=3$~\cite{Qu:2024rns,Nomura:2024atp} and $N=4$~\cite{Ding:2024inn}, and some viable lepton flavor models were constructed. In the present work, we shall consider the level $N=5$ finite modular group $\Gamma_5\cong A_5$ in the framework of non-holomorphic modular flavor symmetry. The three generations of the left-handed (LH) leptons are assigned to an irreducible triplet of $A_5$, and right-handed (RH) leptons are invariant or transform as an irreducible triplet. The non-vanishing polyharmonic Maa{\ss} forms transforming under $\Gamma_N$ require even integer weights. Utilizing the level 5 polyharmonic Maa{\ss} forms with weights $k=\pm4,\, \pm2,\, 0$, we construct all possible lepton models invariant under the $A_5$ modular symmetry. We consider the most economical modular invariant lepton models, no flavon other than $\tau$ is introduced and the light neutrino masses are generated through the Weinberg operator or the type-I seesaw mechanism. Furthermore, the corresponding predictions for lepton mixing parameters and neutrino masses are discussed.

The remaining parts of this paper are organized as follows. We review the concept of modular group and the framework of non-holomorphic modular flavor symmetry in section~\ref{sec:NHMF}. We construct the non-holomorphic lepton flavor models based on the finite modular group $\Gamma_5\cong A_5$ in a systematical way, and the phenomenological predictions for the lepton mixing angles, CP violation phases and neutrino masses are studied in section~\ref{sec:model}.  Two example models are presented and a thorough numerical analysis is performed to discuss their phenomenology in section~\ref{sec:example-models}. Finally we summarize the results and draw the conclusion in section~\ref{sec:conclusion}. The irreducible representations and the related Clebsch-Gordan (CG) coefficients are given in Appendix~\ref{sec:A5_group_theory}, they are indispensable elements when constructing $A_5$ flavor symmetry models.

\section{\label{sec:NHMF}Non-holomorphic modular flavor symmetry}

In this section, we shall give a brief review of the non-holomorphic modular flavor symmetry. The homogeneous modular group $\Gamma \equiv SL(2,\mathbb{Z})$ is the group of two-by-two matrices with integer entries and unit determinant,
\begin{equation}
SL(2,\mathbb{Z})=\left\{
\begin{pmatrix}
a ~& b \\
c ~& d
\end{pmatrix}\bigg| a, b, c, d \in \mathbb{Z},~~ ad -bc = 1     \right\}.
\end{equation}
The modular group can be generated by two matrices
\begin{equation}
S=\begin{pmatrix}
0 ~& 1 \\
-1 ~& 0
\end{pmatrix}, ~~~~
T=\begin{pmatrix}
1 ~& 1 \\
0 ~& 1
\end{pmatrix}\,.
\end{equation}
One can check immediately that $S$ and $T$ satisfy the following relations
\begin{equation}
S^2=-\mathbb{1}\,,~~~S^4=(ST)^3=\mathbb{1},~~~S^2T=TS^2\,,
\end{equation}
which implies $(TS)^3=\mathbb{1}$, where $\mathbb{1}$ refers to the two-dimensional unit matrix. The modular group acts on the complex upper half plane $\mathcal{H}=\left\{\tau\in\mathbb{C}|\Im(\tau)>0\right\}$ by linear fractional transformation
\begin{equation}
\gamma\tau=\frac{a\tau+b}{c\tau+d}\,,~~~~\gamma=\begin{pmatrix}
a ~& b \\
c ~& d
\end{pmatrix}\in SL(2,\mathbb{Z})\,.
\end{equation}
The actions of the generators $S$ and $T$ on the complex modulus $\tau$ read as
\begin{equation}
S\tau=-\frac{1}{\tau}\,,\qquad T\tau=\tau+1\,.
\end{equation}
By applying the modular $T$ transformation in succession, one can map any value of modulus $\tau$ in the upper half plane into the region $-1/2\leq\Re(\tau)\leq1/2$, and it can be further mapped into the fundamental domain,
\begin{equation}
\mathcal{D}=\left\{\tau\in\mathcal{H}\Big| -\frac{1}{2}\leq\Re(\tau)\leq\frac{1}{2},~|\tau|\geq1\right\}\,.
\end{equation}
Each point in the upper half plane can be mapped into the fundamental domain $\mathcal{D}$ by some modular transformation, nevertheless no two points in the interior of $\mathcal{D}$ are related under modular group. Let $N$ be a positive integer, the principal congruence subgroup of level $N$ is the subgroup
\begin{equation}
\Gamma(N)=\left\{
\begin{pmatrix}
a ~& b \\
c ~& d
\end{pmatrix} \in SL(2,\mathbb{Z}),\quad
\begin{pmatrix}
a ~& b \\
c ~& d
\end{pmatrix}\equiv
\begin{pmatrix}
1 & 0 \\
0 & 1
\end{pmatrix}(\text{mod}~ N)\right\}\,.
\end{equation}
Obviously we see $\Gamma(1)=\Gamma$. The quotients $\Gamma_N\equiv \Gamma/\pm\Gamma(N)$ isomorphic to the permutation groups $S_3$, $A_4$, $S_4$ and $A_5$ for $N=2$, 3, 4 and 5 respectively~\cite{deAdelhartToorop:2011re}. Moreover, the quotients $\Gamma'_N \equiv \Gamma/\Gamma(N)$ are double covering groups of $\Gamma_N$~\cite{Liu:2019khw}.

The modular flavor symmetry is originally formulated in the framework of supersymmetry, in which modular symmetry and supersymmetry constrain the Yukawa couplings to be holomorphic modular forms~\cite{Feruglio:2017spp}. Recently the formalism of the non-holomorphic modular flavor symmetry was proposed and supersymmetry is unnecessary in principle~\cite{Qu:2024rns}. The modular invariant Lagrangian is built from matter fields and polyharmonic Maa{\ss} forms of level $N$. The polyharmonic Maa{\ss} forms $Y(\tau)$ of weight $k$ at level $N$ are functions of the complex modulus $\tau$, and their transformation under the modular symmetry is
\begin{equation}
Y(\tau)\mapsto Y(\gamma\tau)=(c\tau+d)^{k}Y(\tau),~~~\gamma=\begin{pmatrix}
a ~& b \\
c ~& d
\end{pmatrix}\in\Gamma(N)\,.
\end{equation}
Moreover, they have to satisfy the following Laplacian condition and proper growth condition~\cite{Qu:2024rns},
\begin{eqnarray}
\nonumber&&~~~~\left(-4y^2\dfrac{\partial}{\partial\tau}\dfrac{\partial}{\partial \bar{\tau}} +2iky\dfrac{\partial}{\partial\bar{\tau}}\right)Y(\tau)=0\,,\\
&&Y(\tau)=\mathcal{O}(y^\alpha) ~~~\text{as $y\rightarrow +\infty$, uniformly in $x$}\,,
\end{eqnarray}
where $\alpha$ is some real parameter and $\tau\equiv x+i y$, and the Laplacian condition can be equivalently expressed as $\partial_{\tau}\left[ y^{k}\partial_{\bar{\tau}} Y(\tau)\right]=0$. The polyharmonic Maa{\ss} form $Y(\tau)$ can contain both holomorphic part and non-holomorphic part, and its Fourier expansion is of the following form
\begin{eqnarray}
\label{eq:poly-harmonic-maass}Y(\tau)=\sum^{+\infty}_{n=0} c^{+}_{n}q^n_N + c^{-}_{0}y^{1-k}+ \sum^{+\infty}_{n=1} c^{-}_{n}\,\Gamma(1-k, 4\pi ny/N)\,q^{-n}_N\,,~~~q_N=e^{2\pi i\tau/N} \,,
\end{eqnarray}
where $\Gamma(s, z)$ is the incomplete gamma function:
\begin{eqnarray}
\label{eq:incomplete-gamma}\Gamma(s,z)=\int_z^{+\infty} e^{-t}t^{s-1}\,dt\,.
\end{eqnarray}
It is straightforward to determine $\Gamma(1, z)=e^{-z}$. For other integer values of $s$, one can obtain the analytical expression of the incomplete gamma function by using the following recursion relation,
\begin{eqnarray}
\label{eq:incomplete-gamma-recursion}\Gamma(s+1,z)=s\Gamma(s,z) + z^s\,e^{-z}\,.
\end{eqnarray}
It is known that the modular weights of modular forms are non-negative integers, while the modular weights of polyharmonic Maa{\ss} forms can be generic integers which can be positive, negative or zero. Note that the product of two polyharmonic Maa{\ss} forms of weights $k$ and $k'$ generally is not a weight $k+k'$ polyharmonic Maa{\ss} form, because the Laplacian condition could be spoiled for $k, k'<0$. The weight $k$ polyharmonic Maa{\ss} forms of level $N$ can be lifted from the weight $2-k$ modular forms of level $N$~\cite{Qu:2024rns}. The non-holomorphic part is vanishing at weights $k\geq3$ while there exist non-holomorphic polyharmonic Maa{\ss} forms at weights $k\leq2$. The weights $k$ polyharmonic Maa{\ss} forms of level $N$ are invariant up to the automorphic factor $(c\tau+d)^k$ under $\Gamma(N)$, nevertheless they transform under the finite modular group $\Gamma^\prime_N$ for generic integer $k$ and $\Gamma_N$ for even $k$. One can choose a basis in which this transformation is described by irreducible representations of $\Gamma^\prime_N$ (or $\Gamma_N$ for even $k$), so that the polyharmonic Maa{\ss} forms of weight $k$ and level $N$ can be arranged into irreducible multiplets of $\Gamma^\prime_N$ (or $\Gamma_N$)~\cite{Qu:2024rns}.

The modular symmetry can strongly constrain the Yukawa couplings and the fermion mass structure. The transformation property of the matter field is fully specified by its modular weight and its transformation under the finite modular group $\Gamma_N$ or $\Gamma^{\prime}_N$. To be more specific, adopting the two-component spinor notation, the Weyl spinors $\psi$, $\psi^c$ and the Higgs field $H$ transform under the modular group as
\begin{eqnarray}
\nonumber&&\psi(x)\rightarrow (c\tau+d)^{-k_{\psi}}\rho_{\psi}(\gamma) \psi(x)\,,\\
\nonumber &&\psi^c(x)\rightarrow (c\tau+d)^{-k_{\psi^c}}\rho_{\psi^c}(\gamma)\psi^c(x)\,,\\
&&H(x)\rightarrow (c\tau+d)^{-k_{H}}\rho_{H}(\gamma)H(x)\,,
\end{eqnarray}
where the modular weights $k_{\psi}$, $k_{\psi^c}$, $k_{H}$ are integers, $\rho_{\psi}(\gamma)$, $\rho_{\psi^c}(\gamma)$ and $\rho_{H}(\gamma)$ are unitary representations of $\Gamma_N$ (or $\Gamma'_N$). Then the Yukawa interaction can be written as
\begin{eqnarray}
\label{eq:Yukawa-MFS}\mathcal{L}_Y=-Y^{(k_Y)}(\tau) \psi^c\psi H+\text{h.c.}\,,
\end{eqnarray}
where the Higgs field $H$ should be its complex conjugate $H^{*}$ for the down-type quarks and charged leptons, and we drop the gauge indices. Requiring modular invariance of the above Lagrangian, the field coupling $Y^{(k_Y)}(\tau)$ should be a multiplet of weight $k_Y$ and level $N$ polyharmonic Maa{\ss} forms, and its modular transformation is given by
\begin{equation}
Y^{(k_Y)}(\tau)\mapsto Y^{(k_Y)}(\gamma\tau)=(c\tau+d)^{k_Y}\rho_{Y}(\gamma)Y^{(k_Y)}(\tau)\,,
\end{equation}
where $\rho_{Y}$ is a representation of $\Gamma'_N$ (or $\Gamma_N$ for even $k_Y$), and the modular weight $k_Y$ and the representation $\rho_{Y}$ should obey the following conditions
\begin{equation}
k_Y=k_{\psi^c}+k_{\psi}+k_H\,, \qquad  \rho_{Y}\otimes\rho_{\psi^c}\otimes\rho_{\psi}\otimes\rho_H \ni\mathbf{1}\,.
\end{equation}
The gCP symmetry can be consistently included in the above framework of non-holomorphic modular symmetry~\cite{Qu:2024rns}. It acts on a field multiplet $\varphi(x)$ as $\varphi(x)\rightarrow X_{\bm{r}}\varphi^{*}(x_{\mathcal{P}})$ with $x_{\mathcal{P}}=(t, -\mathbf{x})$. It turns out that the gCP transformation has the canonical form $X_{\bm{r}}=\mathbb{1}_{\bm{r}}$ in the basis where both modular generators $S$ and $T$ are represented by unitary and symmetric matrices~\cite{Novichkov:2019sqv}. As a result, the gCP invariance enforces the coupling constants accompanying each invariant singlet in the Lagrangian to be real. In the present work, we consider the finite modular group $\Gamma_5\cong A_5$ of level $N=5$. We shall employ the representation basis for the $A_5$ generators $S$ and $T$ shown in Appendix~\ref{sec:A5_group_theory}. One see that $T$ is represented by diagonal matrices in different $A_5$ irreducible representations, and the representation matrices of $S$ are real and symmetric. Hence the gCP symmetry amounts to requiring real couplings in our case. In the framework of non-holomorphic modular flavor symmetry, modular invariance constrains the Yukawa couplings to be polyharmonic Maa{\ss} forms of level $5$. The explicit forms of their Fourier expansion have been presented in Ref.~\cite{Qu:2024rns}. Here we do not show these lengthy expressions and only summarize the polyharmonic Maa{\ss} form multiplets of level $N=5$ and weights $k=\pm4,\,\pm2,\,0$ in table~\ref{tab:MF_summary}.

\begin{table}[t!]
\centering
\begin{tabular}{|c|c|} \hline  \hline
Weight  & Polyharmonic  Maa{\ss} form $Y^{(k)}_{\bm{r}}$   \\  \hline
 &  \\[-0.18in]
$k=-4$ &  $Y_{\bm{1}}^{(-4)}$\,, $Y^{(-4)}_{\bm{3}}$\,, $Y^{(-4)}_{\bm{3^{\prime}}}$\,,  $Y^{(-4)}_{\bm{5}}$ \\
 &    \\[-0.18in] \hline
 & \\[-0.18in]
                
$k=-2$ &  $Y^{(-2)}_{\bm{1}}$\,, $Y^{(-2)}_{\bm{3}}$\,, $Y^{(-2)}_{\bm{3^{\prime}}}$\,, $Y^{(-2)}_{\bm{5}}$   \\
 &   \\[-0.18in] \hline
 &  \\[-0.18in]
                
$k=0$ &   $Y^{(0)}_{\bm{1}}$\,,  $Y^{(0)}_{\bm{3}}$\,, $Y^{(0)}_{\bm{3^{\prime}}}$\,, $Y^{(0)}_{\bm{5}}$  \\
 &   \\[-0.18in] \hline
 &  \\[-0.18in]
                
$k=2$ &  $Y^{(2)}_{\bm{1}}$\,, $Y^{(2)}_{\bm{3}}$\,, $Y^{(2)}_{\bm{3^{\prime}}}$\,, $Y^{(2)}_{\bm{5}}$ \\
 &    \\[-0.18in] \hline
 & \\[-0.18in]
                
$k=4$ &  $Y_{\bm{1}}^{(4)}$\,,  $Y^{(4)}_{\bm{3}}$\,, $Y^{(4)}_{\bm{3^{\prime}}}$\,,  $Y^{(4)}_{\bm{4}}$\,,  $Y^{(4)}_{\bm{5}I}$,  $Y^{(4)}_{\bm{5}II}$ \\ [0.02in] \hline \hline
\end{tabular}
\caption{\label{tab:MF_summary} Polyharmonic Maa{\ss} form multiplets of level $5$ and weights $k=\pm4,\,\pm2,\,0$, where the subscript $\bm{r}$ denotes the transformation property under finite group $A_{5}$. Here $Y^{(k)}_{\bm{5}I}$ and $Y^{(k)}_{\bm{5}II}$ stand for two linear independent quintuplets of Maa{\ss} forms at weight 4. The  explicit forms of these polyharmonic  Maa{\ss} form multiplets can be found in Ref.~\cite{Qu:2024rns}. }
\end{table}

\section{\label{sec:model}General model building}

In this section, we shall perform a systematic classification of all minimal lepton mass models based on the finite modular symmetry $\Gamma_{5}\cong A_{5}$. It is realized from non-supersymmetric case with the polyharmonic  Maa{\ss} forms of level $N=5$ which can be decomposed into the irreducible multiplets of the finite group $A_{5}$~\cite{Qu:2024rns}, as is shown in table~\ref{tab:MF_summary}. In the present work,  neutrinos are assumed to be Majorana particles and their masses are considered to arise from  the Weinberg operator or the type I seesaw mechanism with three RH neutrinos. No flavon other than complex modulus $\tau$ will be introduced, and the modular symmetry is broken when $\tau$ obtains a vacuum expectation value. The Higgs doublet field $H$ is assumed to be $A_{5}$ singlet with vanishing modular weight. We assume that the three generations of LH lepton doublets $L=(L_{1},L_{2},L_{3})^{T}$ with weight $k_{L}$ and the three RH neutrinos $N^c=(N^c_{1},N^c_{2},N^c_{3})^{T}$ with weight $k_{N}$ in the type I seesaw models transform as triplet $\bm{3}$ or $\bm{3^{\prime}}$. The three RH charged leptons $E^{c}_{1,2,3}$ are  assigned to singlet $\bm{1}$ or a triplet $\bm{3}$ or $\bm{3^{\prime}}$ under $A_{5}$ modular group. For the former assignment, they are distinguished by their modular weights $k_{1,2,3}$. For the latter assignment, the modular weight of $E^c=(E^c_{1},E^c_{2},E^c_{3})^{T}$ is denoted as $k_{E}$. For each representation assignment, there are in principle infinitely possible weight assignments for lepton fields, and the number of independent couplings of the charged lepton mass terms and neutrino mass terms generally increases with the weight of the involved modular forms. We shall consider  the polyharmonic  Maa{\ss} forms of even weight $k=-4$ to $k=4$, and  employ potentially the lowest weight  modular forms as much as possible in order to reduce free parameters.

\subsection{\label{sec:ch_mass}Charged lepton sector }

\begin{table}[t!]
        \centering
        \renewcommand{\arraystretch}{1.3}
        \begin{tabular}{c}
                \begin{tabular}{|c|c|c|c|c|c|}  \hline\hline
                        \texttt{Cases} & $C_{1}\,(C^{\prime}_{1})$  & $C_{2}\,(C^{\prime}_{2})$  &  $C_{3}\,(C^{\prime}_{3})$ & $C_{4}\,(C^{\prime}_{4})$ & $C_{5}\,(C^{\prime}_{5})$ \\  \hline
                        $(k_{1}+k_{L},k_{2}+k_{L},k_{3}+k_{L})$& $(-4,-2,0)$ & $(-4,-2,2)$ & $(-4,-2,4)$ &  $(-4,0,2)$ &  $(-4,0,4)$ \\ \hline
                                \texttt{Cases} & $C_{6}\,(C^{\prime}_{6})$  & $C_{7}\,(C^{\prime}_{7})$  &  $C_{8}\,(C^{\prime}_{8})$ & $C_{9}\,(C^{\prime}_{9})$ & $C_{10}\,(C^{\prime}_{10})$ \\  \hline
                        $(k_{1}+k_{L},k_{2}+k_{L},k_{3}+k_{L})$& $(-4,2,4)$ & $(-2,0,2)$ & $(-2,0,4)$ &  $(-2,2,4)$ &  $(0,2,4)$ \\ \hline\hline
                \end{tabular}
        \end{tabular}
\caption{\label{tab:sum_ch}The 20 different possible cases for the  assignments that the LH lepton fields $L$ transform as a triplet under $A_5$ and the RH charged leptons are $A_5$ singlets. The ten cases $C_{i}$ ($C^{\prime}_{i}$) are for assignment $L\sim\bm{3}$ ($L\sim\bm{3^{\prime}}$) and $E^{c}_{1,2,3}\sim\bm{1}$ with different modular weights, and the corresponding charged lepton mass matrices take the form in Eq.~\eqref{eq:ch_mass1} (Eq.~\eqref{eq:ch_mass2}).}
\end{table}

Firstly, let us consider the assignments that the three RH charged leptons are all singlet $\bm{1}$ of $A_{5}$. Then if the LH lepton fields $L$ are embedded into the triplet $\bm{3}$ (or $\bm{3^{\prime}}$), modular form multiplet in the representation $\bm{3}$ (or $\bm{3^{\prime}}$) should be invoked in the charged lepton mass terms. As a consequence, there are only two different assignments for lepton fields and we will list the Lagrangian and charged lepton mass matrices for the two cases in the following

\begin{itemize}[labelindent=-0.8em, leftmargin=1.4em]
        
\item[~~(\romannumeral1)]{$L\sim\bm{3}$ and $E^{c}_{1,2,3}\sim\bm{1}$ }

The Lagrangian for the charged lepton Yukawa coupling reads as
\begin{equation}
\label{eq:Le_1st}-\mathcal{L}_e=\alpha\left(E^{c}_1LY^{(k_{1}+k_{L})}_{\bm{3}}H^{*}\right)_{\bm{1}}+
\beta\left(E^{c}_2LY^{(k_{2}+k_{L})}_{\bm{3}}H^{*}\right)_{\bm{1}}+\gamma\left(E^{c}_3LY^{(k_{3}+k_{L})}_{\bm{3}}H^{*}\right)_{\bm{1}}\,,
\end{equation}
where $Y^{(k_{i}+k_{L})}_{\bm{3}}$ ($i=1,2,3$) stand for level 5 polyharmonic  Maa{\ss} forms transforming as $\bm{3}$ under $A_5$ up to automorphy factor.  Notice that if there are several linearly independent polyharmonic Maa{\ss} form multiplets in the same representation $\bm{3}$  at a given weight $k_{i}+k_{L}$, their contributions  to the  Lagrangian are of similar form which  can be straightforwardly read out from the general results in the following. Guided by the principle of minimality and simplicity, the polyharmonic  Maa{\ss} forms of those weights which only involve one independent triplet polyharmonic  Maa{\ss} form in the representation $\bm{3}$ are imposed in the charged lepton sector. From  table~\ref{tab:MF_summary}, we find  that the possible values of the modular weights  $k_{i}+k_{L}$ are  $\pm4,\,\pm2$ and $0$.  Notice that any two rows of a charged lepton mass matrix should not be proportional to each other, otherwise at least one charged lepton would be massless. Furthermore, the permutations of any two rows of the charged lepton mass matrix amounts a redefinition of the RH charged lepton fields, and thus the predictions for charged lepton masses and lepton mixing matrix are unchanged. Without loss of generality, we take $k_{1}+k_{L}<k_{2}+k_{L}<k_{3}+k_{L}$. In the end, we find there are totally 10 possible charged lepton mass matrices from the 10 possible independent assignments of the values of $k_{i}+k_{L}$. The 10 cases are labeled as $C_{j}$ ($j=1,2,\cdots,10$) and summarized in table~\ref{tab:sum_ch}. The three terms in Eq.~\eqref{eq:Le_1st} have similar form, and they contribute to three rows of the charged lepton mass matrix, respectively.  We can read out the charged lepton mass matrices for the 10 cases as follows
\begin{equation}\label{eq:ch_mass1}
M_e(k_{1}+k_{L},k_{2}+k_{L},k_{3}+k_{L})=\begin{pmatrix}
                \alpha Y^{(k_{1}+k_{L})}_{\bm{3}, 1}  ~&~ \alpha Y^{(k_{1}+k_{L})}_{\bm{3}, 3} ~&~ \alpha Y^{(k_{1}+k_{L})}_{\bm{3}, 2}  \\
                \beta Y^{(k_{2}+k_{L})}_{\bm{3}, 1}  ~&~ \beta Y^{(k_{2}+k_{L})}_{\bm{3}, 3} ~&~ \beta Y^{(k_{2}+k_{L})}_{\bm{3}, 2}  \\
                \gamma Y^{(k_{3}+k_{L})}_{\bm{3}, 1}  ~&~\gamma Y^{(k_{3}+k_{L})}_{\bm{3}, 3} ~&~ \gamma Y^{(k_{3}+k_{L})}_{\bm{3}, 2}
        \end{pmatrix}v\,,
\end{equation}
where the charged lepton mass matrix $M_e$ is given in the right-left basis $E^c\,M_e\,L$ with $ v =\langle H\rangle$, and we denote $Y^{(k_{i}+k_{L})}_{\bm{3}}\equiv(Y^{(k_{i}+k_{L})}_{\bm{3}, 1}, Y^{(k_{i}+k_{L})}_{\bm{3}, 2}, Y^{(k_{i}+k_{L})}_{\bm{3}, 3})^{T}$. For all the 10 cases, the phases of all the three couplings $\alpha$, $\beta$ and $\gamma$ can be absorbed into the RH charged leptons $E^{c}_{1}$, $E^{c}_{2}$ and $E^{c}_{3}$ respectively, and they can be taken to be real.

\item[~~(\romannumeral2)]{$L\sim\bm{3^{\prime}}$ and $E^{c}_{1,2,3}\sim\bm{1}$ }

Similarly, the charged lepton mass terms of the Lagrangian  are
\begin{equation}
-\mathcal{L}_e=\alpha\left(E^{c}_1LY^{(k_{1}+k_{L})}_{\bm{3^{\prime}}}H^{*}\right)_{\bm{1}}+
\beta\left(E^{c}_2LY^{(k_{2}+k_{L})}_{\bm{3^{\prime}}}H^{*}\right)_{\bm{1}}+\gamma\left(E^{c}_3LY^{(k_{3}+k_{L})}_{\bm{3^{\prime}}}H^{*}\right)_{\bm{1}}\,,
\end{equation}
 and the corresponding lepton mass matrix takes the following form
\begin{equation}\label{eq:ch_mass2}
M_e(k_{1}+k_{L},k_{2}+k_{L},k_{3}+k_{L})=\begin{pmatrix}
                \alpha Y^{(k_{1}+k_{L})}_{\bm{3^{\prime}}, 1}  ~&~ \alpha Y^{(k_{1}+k_{L})}_{\bm{3^{\prime}}, 3} ~&~ \alpha Y^{(k_{1}+k_{L})}_{\bm{3^{\prime}}, 2}  \\
                \beta Y^{(k_{2}+k_{L})}_{\bm{3^{\prime}}, 1}  ~&~ \beta Y^{(k_{2}+k_{L})}_{\bm{3^{\prime}}, 3} ~&~ \beta Y^{(k_{2}+k_{L})}_{\bm{3^{\prime}}, 2}  \\
                \gamma Y^{(k_{3}+k_{L})}_{\bm{3^{\prime}}, 1}  ~&~\gamma Y^{(k_{3}+k_{L})}_{\bm{3^{\prime}}, 3} ~&~ \gamma Y^{(k_{3}+k_{L})}_{\bm{3^{\prime}}, 2}
        \end{pmatrix}v\,,
\end{equation}
where $k_{i}+k_{L}=\pm4,\,\pm2,\,0$ with $k_{1}+k_{L}<k_{2}+k_{L}<k_{3}+k_{L}$, and parameters $\alpha$, $\beta$ and $\gamma$ can be taken to be real. There are also 10 independent weight assignments which are labeled as $C^{\prime}_{j}$ ($j=1,2,\cdots,10$) and are given in table~\ref{tab:sum_ch}.

\end{itemize}

In the following, we proceed to consider the assignments that the three RH charged leptons $E^{c}$ also transform as a triplet $\bm{3}$ or $\bm{3^{\prime}}$ under $A_{5}$. Note that $LE^{c}$ decomposes as $\bm{3}\otimes\bm{3}=\bm{1}\oplus\bm{3_{A}}\oplus\bm{5_{S}}$, $\bm{3^{\prime}}\otimes\bm{3^{\prime}}=\bm{1}\oplus\bm{3^{\prime}_{A}}\oplus\bm{5_{S}}$ or $\bm{3}\otimes\bm{3^{\prime}}=\bm{4}\oplus\bm{5}$, using the CG coefficients shown in table~\ref{tab:A5_CG}, we can write out the modular invariant charged lepton mass terms and extract the mass matrix for each assignment as following:

\begin{itemize}[labelindent=-0.8em, leftmargin=1.4em]
        
\item[~~(\romannumeral3)]{$L\sim\bm{3}$ and $E^{c}\sim\bm{3}$ }
        
In this case, the most general Lagrangian $\mathcal{L}_e$  for the charged lepton masses is given by :
\begin{equation}
-\mathcal{L}_e=\alpha\left(E^{c}LY^{(k_{E}+k_{L})}_{\bm{1}}H^{*}\right)_{\bm{1}}+
        \beta \left(E^{c}LY^{(k_{E}+k_{L})}_{\bm{3}}H^{*}\right)_{\bm{1}}+\gamma\left(E^{c}LY^{(k_{E}+k_{L})}_{\bm{5}}H^{*}\right)_{\bm{1}}\,,
\end{equation}
where the coupling $\alpha$ is real, and the  couplings $\beta$ and $\gamma$ are generally complex in the case that gCP symmetry isn't considered.  If the modular group $A_{5}$ is extended to combine with the gCP symmetry, all the couplings are further constrained to be real in our working basis. For the simplest weight assignment $w=k_{E}+k_{L}=-4,\,-2,\,0,\,2$, the charged lepton mass matrix reads as,
\begin{equation}\label{eq:ch_mass_33}
\hskip-0.13in 
M_{e}(w)=\begin{pmatrix}
\alpha  Y^{(w)}_{\bm{1}}+2 \gamma  Y^{(w)}_{\bm{5},1} & -\beta Y^{(w)}_{\bm{3},3}-\sqrt{3} \gamma  Y^{(w)}_{\bm{5},5} & \beta  Y^{(w)}_{\bm{3},2}-\sqrt{3} \gamma  Y^{(w)}_{\bm{5},2} \\
        \beta  Y^{(w)}_{\bm{3},3}-\sqrt{3} \gamma  Y^{(w)}_{\bm{5},5} & \sqrt{6} \gamma  Y^{(w)}_{\bm{5},4} & \alpha  Y^{(w)}_{\bm{1}}-\beta  Y^{(w)}_{\bm{3},1}-\gamma  Y^{(w)}_{\bm{5},1} \\
        -\beta  Y^{(w)}_{\bm{3},2}-\sqrt{3} \gamma  Y^{(w)}_{\bm{5},2} & \alpha  Y^{(w)}_{\bm{1}}+\beta  Y^{(w)}_{\bm{3},1}-\gamma  Y^{(w)}_{\bm{5},1} & \sqrt{6} \gamma  Y^{(w)}_{\bm{5},3} \\
                \end{pmatrix}v\,.
\end{equation}
Note that the above mass matrix involves two more parameters than $C_{j}(C^{\prime}_{j})$ in table~\ref{tab:sum_ch} when gCP symmetry is not imposed. Generally one needs to tune the values of couplings $\alpha$, $\beta$, $\gamma$ to reproduce the charged lepton mass hierarchy for this type of assignment,

\item[~~(\romannumeral4)]{$L\sim\bm{3^{\prime}}$ and $E^{c}\sim\bm{3^{\prime}}$ }
                
Similar to previous case, the charged lepton Lagrangian invariant under the $A_5$ modular symmetry reads as
\begin{equation}
-\mathcal{L}_e=\alpha\left(E^{c}LY^{(k_{E}+k_{L})}_{\bm{1}}H^{*}\right)_{\bm{1}}+
        \beta \left(E^{c}LY^{(k_{E}+k_{L})}_{\bm{3^{\prime}}}H^{*}\right)_{\bm{1}}+\gamma\left(E^{c}LY^{(k_{E}+k_{L})}_{\bm{5}}H^{*}\right)_{\bm{1}}\,,
\end{equation}
which leads to the following charged lepton  mass matrix
\begin{equation}\label{eq:ch_mass_3p3p}
\hskip-0.13in
M_{e}(w)=
\begin{pmatrix}
        \alpha  Y^{(w)}_{\bm{1}}+2 \gamma  Y^{(w)}_{\bm{5},1} & -\beta  Y^{(w)}_{\bm{3^{\prime}},3}-\sqrt{3} \gamma  Y^{(w)}_{\bm{5},4} & \beta  Y^{(w)}_{\bm{3^{\prime}},2}-\sqrt{3} \gamma  Y^{(w)}_{\bm{5},3} \\
        \beta  Y^{(w)}_{\bm{3^{\prime}},3}-\sqrt{3} \gamma  Y^{(w)}_{\bm{5},4} & \sqrt{6} \gamma  Y^{(w)}_{\bm{5},2} & \alpha  Y^{(w)}_{\bm{1}}-\beta  Y^{(w)}_{\bm{3^{\prime}},1}-\gamma  Y^{(w)}_{\bm{5},1} \\
        -\beta  Y^{(w)}_{\bm{3^{\prime}},2}-\sqrt{3} \gamma  Y^{(w)}_{\bm{5},3} & \alpha  Y^{(w)}_{\bm{1}}+\beta  Y^{(w)}_{\bm{3^{\prime}},1}-\gamma  Y^{(w)}_{\bm{5},1} & \sqrt{6} \gamma  Y^{(w)}_{\bm{5},5} \\
                        \end{pmatrix}v\,,
\end{equation}
with the modular weights  $w=k_{E}+k_{L}$.

\item[~~(\romannumeral5)]{$L\sim\bm{3}$ and $E^{c}\sim\bm{3^{\prime}}$ }

From the multiplication rule $\bm{3}\otimes\bm{3^{\prime}}=\bm{4}\oplus\bm{5}$, we find that the modular invariant Lagrangian $\mathcal{L}_e$ for the charged lepton masses can be written as :
\begin{equation}
-\mathcal{L}_e=\alpha\left(E^{c}LY^{(k_{E}+k_{L})}_{\bm{5}}H^{*}\right)_{\bm{1}}+\beta\left(E^{c}LY^{(k_{E}+k_{L})}_{\bm{4}}H^{*}\right)_{\bm{1}}\,.
\end{equation}
 When the modular weight of $w=k_{E}+k_{L}$ is taken to be $-4$, $-2$, $0$ and $2$, the charged lepton mass matrix is given by
\begin{equation}
M_{e}(w)=\begin{pmatrix}
        \sqrt{3} \alpha  Y^{(w)}_{\bm{5},1} & \alpha  Y^{(w)}_{\bm{5},5} &\alpha  Y^{(w)}_{\bm{5},2} \\
        \alpha  Y^{(w)}_{\bm{5},4} & -\sqrt{2} \alpha  Y^{(w)}_{\bm{5},3} & -\sqrt{2} \alpha  Y^{(w)}_{\bm{5},5} \\
        \alpha  Y^{(w)}_{\bm{5},3} & -\sqrt{2} \alpha  Y^{(w)}_{\bm{5},2} & -\sqrt{2} \alpha  Y^{(w)}_{\bm{5},4} \\
\end{pmatrix}v\,,
\end{equation}
which only involves one real input parameter $\alpha$. The reason is that there is no quartet modular form at these modular weights. It turns out to be very difficult to accommodate the three charged lepton mass with only one coupling $\alpha$. For $k_{E}+k_{L}=4$, the charged lepton mass matrix is given by
\begin{eqnarray}
\nonumber 
M_{e}&=&\left[\begin{pmatrix}
\sqrt{3} \alpha _{1} Y^{(4)}_{\bm{5}I,1}+       \sqrt{3} \alpha _{2} Y^{(4)}_{\bm{5}II,1}  & \alpha _{1} Y^{(4)}_{\bm{5}I,5}+\alpha _{2} Y^{(4)}_{\bm{5}II,5} &\alpha _{1} Y^{(4)}_{\bm{5}I,2}+\alpha _{2} Y^{(4)}_{\bm{5}II,2} \\
\alpha _{1} Y^{(4)}_{\bm{5}I,4}+\alpha _{2} Y^{(4)}_{\bm{5}II,4} & -\sqrt{2} \alpha _{1} Y^{(4)}_{\bm{5}I,3}-\sqrt{2} \alpha _{2} Y^{(4)}_{\bm{5}II,3} & -\sqrt{2} \alpha _{1} Y^{(4)}_{\bm{5}I,5}-\sqrt{2} \alpha _{2} Y^{(4)}_{\bm{5}II,5} \\
\alpha _{1} Y^{(4)}_{\bm{5}I,3}+\alpha _{2} Y^{(4)}_{\bm{5}II,3} & -\sqrt{2} \alpha _{1} Y^{(4)}_{\bm{5}I,2}-\sqrt{2} \alpha _{2} Y^{(4)}_{\bm{5}II,2}  & -\sqrt{2} \alpha _{1} Y^{(4)}_{\bm{5}I,4}-\sqrt{2} \alpha _{2} Y^{(4)}_{\bm{5}II,4} \\
        \end{pmatrix} \right. \\
        &&\left.+\beta\begin{pmatrix}
                0  & \sqrt{2} Y^{(4)}_{\bm{4},4} & \sqrt{2}   Y^{(4)}_{\bm{4},1}\\
                -\sqrt{2}   Y^{(4)}_{\bm{4},3} & - Y^{(4)}_{\bm{4},2} &  Y^{(4)}_{\bm{4},4} \\
                -\sqrt{2}  Y^{(4)}_{\bm{4},2} &  Y^{(4)}_{\bm{4},1} & - Y^{(4)}_{\bm{4},3} \\
        \end{pmatrix}\right]v\,,
\end{eqnarray}
where parameter $\alpha_{1}$ can taken to be real, while parameters $\alpha_{2}$ and $\beta$ are generally complex. This mass matrix also involves more input parameters than the mass matrices of $C_{i}(C^{\prime}_{i})$ in table~\ref{tab:sum_ch}.

\item[~~(\romannumeral6)]{$L\sim\bm{3^{\prime}}$ and $E^{c}\sim\bm{3}$ }

For this assignment, the charged lepton mass matrices are the transpose of those in case (\romannumeral5).

\end{itemize}

In general, the viable charged lepton mass matrices for triplet RH charged lepton assignment involve more free parameters than those for three singlet RH charged leptons when gCP symmetry is not imposed. Hence we shall analyze the scenario that the three RH charged leptons are $A_5$ singlets in the following.

\subsection{Neutrino sector}

In the present work, the  neutrinos are assumed to be Majorana particles. The light neutrino masses are described by the effective Weinberg operator or arise from the type I seesaw mechanism.  Let us first consider that the general form of the modular invariant Majorana neutrino mass terms described by Weinberg operator. Under the assumption that the LH lepton doublets $L$ transform as a triplet $\bm{3}$ or $\bm{3^{\prime}}$ under $A_{5}$, and the anti-symmetric contractions $(LL)_{\bm{3_{A}}}$ for $L\sim \bm{3}$ and $(LL)_{\bm{3^{\prime}_{A}}}$ for $L\sim \bm{3^{\prime}}$ are vanishing. Thus the singlet polyharmonic Maa{\ss} form $Y^{(2k_{L})}_{\bm{1}}$ or the quintuplet Maa{\ss} form $Y^{(2k_{L})}_{\bm{5}}$ is necessary to obtain nonzero neutrino masses. Then the most general modular invariant Lagrangian for neutrino masses can be written as
\begin{equation}\label{eq:weinberg}
-\mathcal{L}_{\nu}=\frac{g_{1}}{2\Lambda}\left(LLHHY^{(2k_{L})}_{\bm{1}}\right)_{\bm{1}}+\frac{g_{2}}{2\Lambda}\left(LLHHY^{(2k_{L})}_{\bm{5}}\right)_{\bm{1}}\,.
\end{equation}
For the sake of simplicity, we are concerned with the modular weight $k_{L}=-2,-1,0,1$. Then there are 8 independent cases differing in the representation assignment and weight of $L$, and they are labelled as $W_{i}$ and $W^{\prime}_{i}$ ($i=1,2,3,4$) for $L\sim \bm{3}$ and  $L\sim \bm{3^{\prime}}$ respectively, as summarized in table~\ref{tab:sum_WO_SeeSaw}. Applying the decomposition rules of the finite modular group $A_{5}$ given in table~\ref{tab:A5_CG}, we find that $\mathcal{L}_{\nu}$ in  Eq.~\eqref{eq:weinberg} gives rise to the following light neutrino mass matrices
{\small \begin{eqnarray}
\nonumber && \hskip-0.3in 
M_{\nu}(k_{L})=\frac{v^2}{2\Lambda}\left(
\begin{array}{ccc}
        g_{1}Y^{(2k_{L})}_{\bm{1}}+2 g_{2}Y^{(2k_{L})}_{\bm{5},1}& -\sqrt{3} g_{2}Y^{(2k_{L})}_{\bm{5},5}& -\sqrt{3} g_{2}Y^{(2k_{L})}_{\bm{5},2}\\
        -\sqrt{3} g_{2}Y^{(2k_{L})}_{\bm{5},5}& \sqrt{6} g_{2}Y^{(2k_{L})}_{\bm{5},4}& g_{1}Y^{(2k_{L})}_{\bm{1}}-g_{2}Y^{(2k_{L})}_{\bm{5},1}\\
        -\sqrt{3} g_{2}Y^{(2k_{L})}_{\bm{5},2}& g_{1}Y^{(2k_{L})}_{\bm{1}}-g_{2}Y^{(2k_{L})}_{\bm{5},1}& \sqrt{6} g_{2}Y^{(2k_{L})}_{\bm{5},3}\\
\end{array}
\right), \quad \text{for} \quad L\sim \bm{3}\,, \\
\label{eq:WO_mass}\nonumber   &&\hskip-0.3in 
M_{\nu}(k_{L})=\frac{v^2}{2\Lambda}\left(
\begin{array}{ccc}
        g_{1}Y^{(2k_{L})}_{\bm{1}}+2 g_{2}Y^{(2k_{L})}_{\bm{5},1} & -\sqrt{3} g_{2}Y^{(2k_{L})}_{\bm{5},4} & -\sqrt{3} g_{2}Y^{(2k_{L})}_{\bm{5},3} \\
        -\sqrt{3} g_{2}Y^{(2k_{L})}_{\bm{5},4} & \sqrt{6} g_{2}Y^{(2k_{L})}_{\bm{5},2} & g_{1}Y^{(2k_{L})}_{\bm{1}}-g_{2}Y^{(2k_{L})}_{\bm{5},1} \\
        -\sqrt{3} g_{2}Y^{(2k_{L})}_{\bm{5},3} & g_{1}Y^{(2k_{L})}_{\bm{1}}-g_{2}Y^{(2k_{L})}_{\bm{5},1} & \sqrt{6} g_{2}Y^{(2k_{L})}_{\bm{5},5} \\
\end{array}
\right), \quad \text{for} \quad L\sim \bm{3^{\prime}}\,, \\
\end{eqnarray}}
where the phase of parameter $g_{1}$ can be set to be zero, while the phase of $g_{2}$ can not be removed by a field redefinition. Both $g_{1}$ and $g_{2}$ would be real if gCP symmetry is included.

\begin{table}[t!]
\centering
\renewcommand{\tabcolsep}{2.1mm}
\renewcommand{\arraystretch}{1.3}
\begin{tabular}{c}
\begin{tabular}{|c|c|c|c|c|c|c|c|c|c|c|c|}  \hline\hline
\multicolumn{5}{|c|}{ Weinberg operator} \\ \hline
\texttt{Cases} & $W_{1}(W^{\prime}_{1})$  & $W_{2}(W^{\prime}_{2})$ & $W_{3}(W^{\prime}_{3})$ &  $W_{4}(W^{\prime}_{4})$  \\  \hline
$k_{L}$& $-2$ & $-1$ & $0$ &  $1$ \\ \hline \hline
\multicolumn{5}{|c|}{Seesaw } \\ \hline
\texttt{Cases} & $D_{1}\, (D^{\prime}_{1})$  & $D_{2}\, (D^{\prime}_{2})$ & $D_{3}\, (D^{\prime}_{3})$&  $D_{4}\, (D^{\prime}_{4})$ \\  \hline
$k_{L}+k_{N}$& $-4$ & $-2$ & $0$ & $2$  \\ \hline
                        
\texttt{Cases} & $N_{1}(N^{\prime}_{1})$  & $N_{2}(N^{\prime}_{2})$ & $N_{3}(N^{\prime}_{3})$ &  $N_{4}(N^{\prime}_{4})$  \\  \hline
$k_{N}$& $-2$ & $-1$ & $0$ &  $1$ \\ \hline \hline
                        
\end{tabular}
\end{tabular}
\caption{\label{tab:sum_WO_SeeSaw} The values of the modular weights $k_L$ and $k_N$ of the LH lepton $L$ and RH neutrinos $N^c$. The cases $W_{i}(W^{\prime}_{i})$ are for the assignment $L\sim\bm{3}$ ($\bm{3^{\prime}}$) with different value of $k_{L}$, and the light neutrino masses are described by the Weinberg operator. The Majorana mass terms of $N_{i}(N^{\prime}_{i})$ are for the assignment $N^{c}\sim\bm{3}$ ($\bm{3^{\prime}}$) with different weight $k_{N}$, and the Dirac mass terms of $D_{i}\, (D^{\prime}_{i})$  correspond to the assignments $(L,N^{c})\sim(\bm{3},\bm{3^{\prime}})$ ($(L,N^{c})\sim(\bm{3^{\prime}},\bm{3})$) with different weight $k_{L}+k_{N}$. The light neutrino mass matrices of $W_{i}(W^{\prime}_{i})$ are given in Eq.~\eqref{eq:WO_mass}. The Majorana mass matrices of cases $N_{i}(N^{\prime}_{i})$ can be obtained from $M_{N}$ in Eqs.~(\ref{eq:Majorana1_mass}, \ref{eq:Majorana2_mass}). The Dirac mass matrices for cases  $D_{i}\, (D^{\prime}_{i})$  are shown in Eq.~\eqref{eq:Dirac_mass_1} . }
\end{table}

When the neutrino masses are generated by the type I seesaw mechanism, the three RH neutrinos $N^{c}$ are assigned to triplet $\bm{3}$ or $\bm{3^{\prime}}$ of $A_{5}$. Then the  Lagrangian for the neutrino masses can be generally written as
\begin{equation}
-\mathcal{L}_{\nu}=\sum_{\bm{r_{1}}} g\left(N^cLY^{(k_{L}+k_{N})}_{\bm{r_{1}}}H\right)_{\mathbf{1}}+\sum_{\bm{r_{2}}} \frac{h}{2}\Lambda\left(N^cN^cY^{(2k_{N})}_{\bm{r_{2}}}\right)_{\mathbf{1}}\,,
\end{equation}
where the polyharmonic Maa{\ss} form multiplets $Y^{(k_{L}+k_{N})}_{\bm{r_{1}}}$ and $Y^{(2k_{N})}_{\bm{r_{2}}}$ are required to ensure modular invariance, and the explicit forms of them are determined by the weights and representation assignments of $L$ and $N^c$. Similar to the Weinberg operator in Eq.~\eqref{eq:weinberg}, the representation $\bm{r_{2}}$ can be $\bm{1}$ and $\bm{5}$ of $A_5$. We are concerned with the models with minimal number of parameters, consequently we focus on the low weights $k_{N}=-2,-1,0,1$. Then the 8 possible assignments of $N^{c}$ and its weight are labeled as $N_{i}$ and $N^{\prime}_{i}$ ($i=1,2,3,4$) which are summarized in table~\ref{tab:sum_WO_SeeSaw}, and  the corresponding mass matrices for the Majorana neutrinos $N^c$ are given by
\begin{eqnarray}
\label{eq:Majorana1_mass} &&\hskip-0.55in
M_{N}(k_{N})=\frac{\Lambda}{2}\left(
\begin{array}{ccc}
h_{1}Y^{(2k_{L})}_{\bm{1}}+2 h_{2}Y^{(2k_{N})}_{\bm{5},1}& -\sqrt{3} h_{2}Y^{(2k_{N})}_{\bm{5},5}& -\sqrt{3} h_{2}Y^{(2k_{N})}_{\bm{5},2}\\
-\sqrt{3} h_{2}Y^{(2k_{N})}_{\bm{5},5}& \sqrt{6} h_{2}Y^{(2k_{N})}_{\bm{5},4}& h_{1}Y^{(2k_{N})}_{\bm{1}}-h_{2}Y^{(2k_{N})}_{\bm{5},1}\\
-\sqrt{3} h_{2}Y^{(2k_{N})}_{\bm{5},2}& h_{1}Y^{(2k_{N})}_{\bm{1}}-h_{2}Y^{(2k_{N})}_{\bm{5},1}& \sqrt{6} h_{2}Y^{(2k_{N})}_{\bm{5},3}\\
\end{array}
\right) \,,
\end{eqnarray}
for $N^{c}\sim \bm{3}$ and
\begin{eqnarray}
\label{eq:Majorana2_mass} && \hskip-0.55in
M_{N}(k_{N})=\frac{\Lambda}{2}\left(
\begin{array}{ccc}
h_{1}Y^{(2k_{N})}_{\bm{1}}+2 h_{2}Y^{(2k_{N})}_{\bm{5},1} & -\sqrt{3} h_{2}Y^{(2k_{N})}_{\bm{5},4} & -\sqrt{3} h_{2}Y^{(2k_{N})}_{\bm{5},3} \\
-\sqrt{3} h_{2}Y^{(2k_{N})}_{\bm{5},4} & \sqrt{6} h_{2}Y^{(2k_{N})}_{\bm{5},2} & h_{1}Y^{(2k_{N})}_{\bm{1}}-h_{2}Y^{(2k_{N})}_{\bm{5},1} \\
-\sqrt{3} h_{2}Y^{(2k_{N})}_{\bm{5},3} & h_{1}Y^{(2k_{N})}_{\bm{1}}-h_{2}Y^{(2k_{N})}_{\bm{5},1} & \sqrt{6} h_{2}Y^{(2k_{N})}_{\bm{5},5} \\
\end{array}
\right)\,.
\end{eqnarray}
for $N^{c}\sim \bm{3^{\prime}}$. 

The Dirac neutrino mass term  is similar to the charged lepton mass terms discussed in cases (\romannumeral3) to (\romannumeral6) of section~\ref{sec:ch_mass}. Firstly, let us consider the assignments that $L$ and $N^{c}$ transform as different triplets of $A_{5}$, then $Y^{(k_{L}+k_{N})}_{\bm{r_{1}}}$ must be quartet $\bm{4}$ or quintuplet $\bm{5}$. When the modular weight $w=k_{L}+k_{N}$ takes the values of $-4$, $-2$, $0$ and $2$, only the quintuplet Maa{\ss} form $Y^{(k_{L}+k_{N})}_{\bm{5}}$ contributes to the Dirac neutrino mass matrix reading as
\begin{eqnarray}
\nonumber M_D(w)&=&\left(
\begin{array}{ccc}
\sqrt{3} gY^{(w)}_{\bm{5},1} & g Y^{(w)}_{\bm{5},5} & gY^{(w)}_{\bm{5},2}\\
gY^{(w)}_{\bm{5},4}& -\sqrt{2} gY^{(w)}_{\bm{5},3} & -\sqrt{2} gY^{(w)}_{\bm{5},5} \\
gY^{(w)}_{\bm{5},3} & -\sqrt{2} gY^{(w)}_{\bm{5},2} & -\sqrt{2} gY^{(w)}_{\bm{5},4}\\
\end{array}
\right)v\,, \quad \text{for}\quad (L,N^{c})\sim(\bm{3},\bm{3^{\prime}})\,, \\
\label{eq:Dirac_mass_1}
M_D(w)&=&\left(
\begin{array}{ccc}
\sqrt{3} gY^{(w)}_{\bm{5},1} & gY^{(w)}_{\bm{5},4} & gY^{(w)}_{\bm{5},3}\\
gY^{(w)}_{\bm{5},5}& -\sqrt{2} gY^{(w)}_{\bm{5},3} & -\sqrt{2} gY^{(w)}_{\bm{5},2} \\
gY^{(w)}_{\bm{5},2} & -\sqrt{2} gY^{(w)}_{\bm{5},5} & -\sqrt{2} gY^{(w)}_{\bm{5},4}\\
\end{array}
\right)v\,, \quad \text{for}\quad (L,N^{c})\sim(\bm{3^{\prime}},\bm{3})\,,
\end{eqnarray}
which depends on a unique overall coupling $g$. The 8 possible Dirac neutrino mass matrices in Eq.~\eqref{eq:Dirac_mass_1} correspond to  the  8 possible independent cases $D_{i}$ and $D^{\prime}_{i}$ listed in table~\ref{tab:sum_WO_SeeSaw}.

If both $L$ and $N^{c}$ are assigned to a same triplet $\bm{3}$ (or $\bm{3^{\prime}}$) of $A_{5}$, the polyharmonic Maa{\ss} form  $Y^{(k_{L}+k_{N})}_{\bm{r_{1}}}(\tau)$ can be $A_5$ singlet $\bm{1}$, triplet $\bm{3}$ (or $\bm{3^{\prime}}$) and quintuplet $\bm{5}$. Since all these three modular multiplets are present at every modular weight, the Dirac neutrino mass matrix involves at least three couplings and the corresponding Dirac neutrino mass matrix is given by
{\small\begin{eqnarray}
\label{eq:Dirac1_mass_2} \hskip-0.3in 
M_{D}(w)&=&
\begin{pmatrix}
g_{1}  Y^{(w)}_{\bm{1}}+2 g_{3}  Y^{(w)}_{\bm{5},1} & -g_{2} Y^{(w)}_{\bm{3},3}-\sqrt{3} g_{3}  Y^{(w)}_{\bm{5},5} & g_{2}  Y^{(w)}_{\bm{3},2}-\sqrt{3} g_{3}  Y^{(w)}_{\bm{5},2} \\
g_{2}  Y^{(w)}_{\bm{3},3}-\sqrt{3} g_{3}  Y^{(w)}_{\bm{5},5} & \sqrt{6} g_{3}  Y^{(w)}_{\bm{5},4} & g_{1}  Y^{(w)}_{\bm{1}}-g_{2}  Y^{(w)}_{\bm{3},1}-g_{3}  Y^{(w)}_{\bm{5},1} \\
-g_{2}  Y^{(w)}_{\bm{3},2}-\sqrt{3} g_{3}  Y^{(w)}_{\bm{5},2} & g_{1}  Y^{(w)}_{\bm{1}}+g_{2}  Y^{(w)}_{\bm{3},1}-g_{3}  Y^{(w)}_{\bm{5},1} & \sqrt{6} g_{3}  Y^{(w)}_{\bm{5},3} \\
\end{pmatrix}v\,,
\end{eqnarray}
for $L,N^{c}\sim\bm{3}$ and
\begin{eqnarray}
\label{eq:Dirac2_mass_2} \hskip-0.3in
M_{D}(w)&=&\begin{pmatrix}
g_{1} Y^{(w)}_{\bm{1}}+2 g_{3}  Y^{(w)}_{\bm{5},1} & -g_{2}  Y^{(w)}_{\bm{3^{\prime}},3}-\sqrt{3} g_{3}  Y^{(w)}_{\bm{5},4} & g_{2}  Y^{(w)}_{\bm{3^{\prime}},2}-\sqrt{3} g_{3}  Y^{(w)}_{\bm{5},3} \\
g_{2} Y^{(w)}_{\bm{3^{\prime}},3}-\sqrt{3} g_{3}  Y^{(w)}_{\bm{5},4} & \sqrt{6} g_{3}  Y^{(w)}_{\bm{5},2} & g_{1}  Y^{(w)}_{\bm{1}}-g_{2}  Y^{(w)}_{\bm{3^{\prime}},1}-g_{3}  Y^{(w)}_{\bm{5},1} \\
-g_{2}  Y^{(w)}_{\bm{3^{\prime}},2}-\sqrt{3} g_{3}  Y^{(w)}_{\bm{5},3} & g_{1}  Y^{(w)}_{\bm{1}}+g_{2}  Y^{(w)}_{\bm{3^{\prime}},1}-g_{3}  Y^{(w)}_{\bm{5},1} & \sqrt{6} g_{3}  Y^{(w)}_{\bm{5},5} \\
\end{pmatrix}v\,.
\end{eqnarray}}
for $L,N^{c}\sim\bm{3^{\prime}}$ with $w=k_{L}+k_{N}=-4,-2,0,2$, respectively. We shall not consider these cases in the following, since we are only interested in the models with less free parameters.

In short, we have obtained 8 minimal models labeled as $W_{i}$ and $W^{\prime}_{i}$ when neutrino masses are generated by the Weinberg operator, and 32 minimal models $D_{i}-N^{\prime}_{j}$ and $D^{\prime}_{i}-N_{j}$ with $i,j=1,2,3,4$ are obtained if neutrino masses are generated by seesaw mechanism. The values of the modular weights $k_L$ and $k_N$ for each case are listed in table~\ref{tab:sum_WO_SeeSaw}. All these resulting light neutrino mass matrices contain 3 real input parameters besides the complex modulus $\tau$ in the case of without gCP symmetry. One more real parameter would be reduced if gCP symmetry is imposed.

\subsection{Numerical results}

In short, taking into account the possible structures of the models with the charged lepton mass matrices summarized in table~\ref{tab:sum_ch}  and neutrino mass matrices summarized  in table~\ref{tab:sum_WO_SeeSaw}, we find that there are a total of 400 minimal non-supersymmetric lepton models based on the finite modular symmetry $A_{5}$: 80 different models $C_{i}-W_{j}$ and $C^{\prime}_{i}-W^{\prime}_{j}$ in which the neutrino masses are described by the effective Weinberg operator,  and 320 different models  $C_{i}-D_{j}-N^{\prime}_{k}$ and $C^{\prime}_{i}-D^{\prime}_{j}-N_{k}$ in which neutrinos gain masses via type I seesaw mechanism, where $i=1,2,\cdots,10$ and $j,k=1,2,3,4$. In each model,  the modular weight assignments of the matter fields can be fixed uniquely.  Here we do not show them.  It is notable that coupling constants $\alpha$, $\beta$ and $\gamma$ in the charged lepton mass matrices can be taken to be positive. If the neutrino masses are generated by  the effective Weinberg operator, the light neutrino mass matrices have two independent real parameters $\left|g_{2}/g_{1}\right|$, $\text{arg}(g_{2}/g_{1})$ besides the overall factor $g_{1}v^2/\Lambda$ and the modulus $\tau$. If the light neutrino masses originate from the type I seesaw mechanism, the light neutrino mass matrices for all models are determined by $\left|h_{2}/h_{1}\right|$ and $\text{arg}(h_{2}/h_{1})$ up to an overall factor $g^2v^2/(h_{1}\Lambda)$ and the modulus $\tau$. Hence the number of independent real free parameters of all the 400 models is eight in the case without gCP symmetry. If gCP symmetry compatible with $A_{5}$ is imposed, all couplings are further constrained to be real in our working basis and the number of free parameters reduces to seven.  

\begin{table}[t!]
\centering
\renewcommand{\tabcolsep}{1.6mm}
\renewcommand{\arraystretch}{1.2}
\begin{tabular}{|c|c|c||c|c|c|c|c|}
\hline \hline
Observable  &  $\text{bf}\pm1\sigma$   & $3\sigma$ region  &   Observable &  $\text{bf}\pm1\sigma$  & $3\sigma$ region     \\ \hline

$\sin^2\theta_{13}$ & $0.02215^{+0.00056}_{-0.00058}$ & $[0.02030,0.02388]$ & $\frac{\Delta m^2_{31}}{10^{-3}\text{eV}^2}$ & $2.513^{+0.021}_{-0.019}$ & $[2.451,2.578]$ \\ [0.050in]

$\sin^2\theta_{12}$ & $0.308^{+0.012}_{-0.011}$ & $[0.275,0.345]$ &  $\frac{\Delta m^2_{21}}{\Delta m^2_{31}}$ & $0.0298^{+0.00079}_{-0.00079}$  & $[0.0268,0.0328]$ \\ [0.050in]

$\sin^2\theta_{23}$  & $0.470^{+0.017}_{-0.013}$  & $[0.435,0.585]$ & $m_e/m_{\mu}$ & $0.004737$ & --- \\ [0.050in]

$\delta_{CP}/\pi$  & $1.178^{+0.144}_{-0.228}$ & $[0.689,2.022]$ &  $m_{\mu}/m_{\tau}$ & $0.05882$ & --- \\ [0.050in]

$\frac{\Delta m^2_{21}}{10^{-5}\text{eV}^2}$ & $7.49^{+0.19}_{-0.19}$ & $[6.92,8.05]$ & $m_{e}/\text{MeV}$ & $0.469652$  &  ---  \\  [0.050in] \hline  \hline

\end{tabular}
\caption{\label{tab:bf_13sigma_data} The global best fit values, $1\sigma$ ranges and $3\sigma$ ranges for mixing parameters and lepton mass ratios, where the experimental data and errors of the lepton mixing parameters and neutrino masses for NO neutrino mass spectrum are obtained from NuFIT 6.0 with Super-Kamiokande atmospheric data~\cite{Esteban:2020cvm}. The $1\sigma$ error of the charged lepton mass ratios are taken to be $0.1\%$ of their central values in $\chi^2$ function.  }
\end{table}

In the following, we shall perform a comprehensive numerical analysis for all the 400 models both with and without gCP symmetry. In order to quantitatively assess how well a model can describe  the current experiment data, we define a $\chi^2$ function to estimate the goodness-of-fit of a set of chosen values of the input parameters,
\begin{equation}\label{eq:chisq}
        \chi^2 = \sum_{i=1}^7 \left( \frac{P_i-O_i}{\sigma_i}\right)^2\,,
\end{equation}
where $O_{i}$ and $\sigma_{i}$ represent the global best values and  $1\sigma$ deviations respectively of  the following seven dimensionless observable quantities for normal ordering (NO) neutrino masses :
\begin{equation}\label{eq:obs_qua}
        m_e/m_\mu, ~~ m_\mu/m_\tau,~~\sin^2\theta_{12},~~ \sin^2\theta_{13},~~ \sin^2\theta_{23}, ~~\delta_{CP},~~\Delta m^2_{21}/\Delta m^2_{31}\,,
\end{equation}
where the contribution of the Dirac phase $\delta_{CP}$ is also included in the $\chi^2$ function. The global best fit values and $1\sigma$ uncertainties are summarized in table~\ref{tab:bf_13sigma_data}. $P_i$ in Eq.~\eqref{eq:chisq}  are the theoretical predictions for the above seven physical observable quantities, and they depend on the values of the following six dimensionless input parameters
\begin{equation}\label{eq:six_inputs}
\Re{\tau},  \quad \Im{\tau}, \quad \beta /\alpha, \quad \gamma /\alpha, \quad
|g_2/g_1| ~\left(\text{or}~|h_{2}/h_{1}|\right),\quad \text{arg}{(g_2/g_1)}~\left(\text{or}~\text{arg}{(h_2/h_1)}\right)\,,
\end{equation}
for Weinberg operator (seesaw) models. When gCP symmetry is included, the phase parameters  $\text{arg}{(g_2/g_1)}$ and $\text{arg}{(h_2/h_1)}$ are equal to $0$ or $\pi$. Note that the overall parameters $\alpha v$ of charged lepton mass matrices and $g_{1}v^2/\Lambda$ (or $g^2v^2/(h_{1}\Lambda)$) of light neutrino mass matrices are determined by the measured electron mass and the solar neutrino mass square difference $\Delta m^{2}_{21}$, respectively.

\begin{figure}[t!]
\centering
\begin{tabular}{c}
\includegraphics[width=0.9\linewidth]{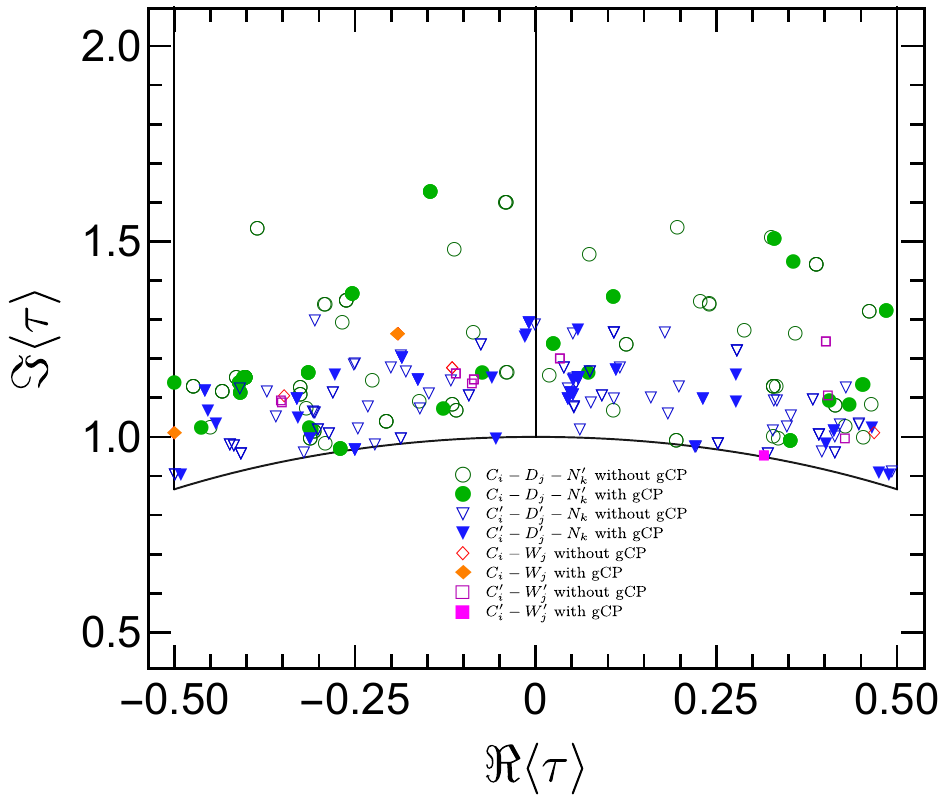}
\end{tabular}
\caption{\label{fig:bf_tau}  The best fit values of $\tau$ for the 21 (4) viable Weinberg operator  models $C_{i}-W_{j}$ and $C^{\prime}_{i}-W^{\prime}_{j}$,  and  the 174 (100) viable seesaw models $C_{i}-D_{j}-N^{\prime}_{k}$ and $C^{\prime}_{i}-D^{\prime}_{j}-N_{k}$ in the case of without (with) gCP symmetry. }
\end{figure}

\begin{figure}[t!]
\centering
\begin{tabular}{c}
\includegraphics[width=0.9\linewidth]{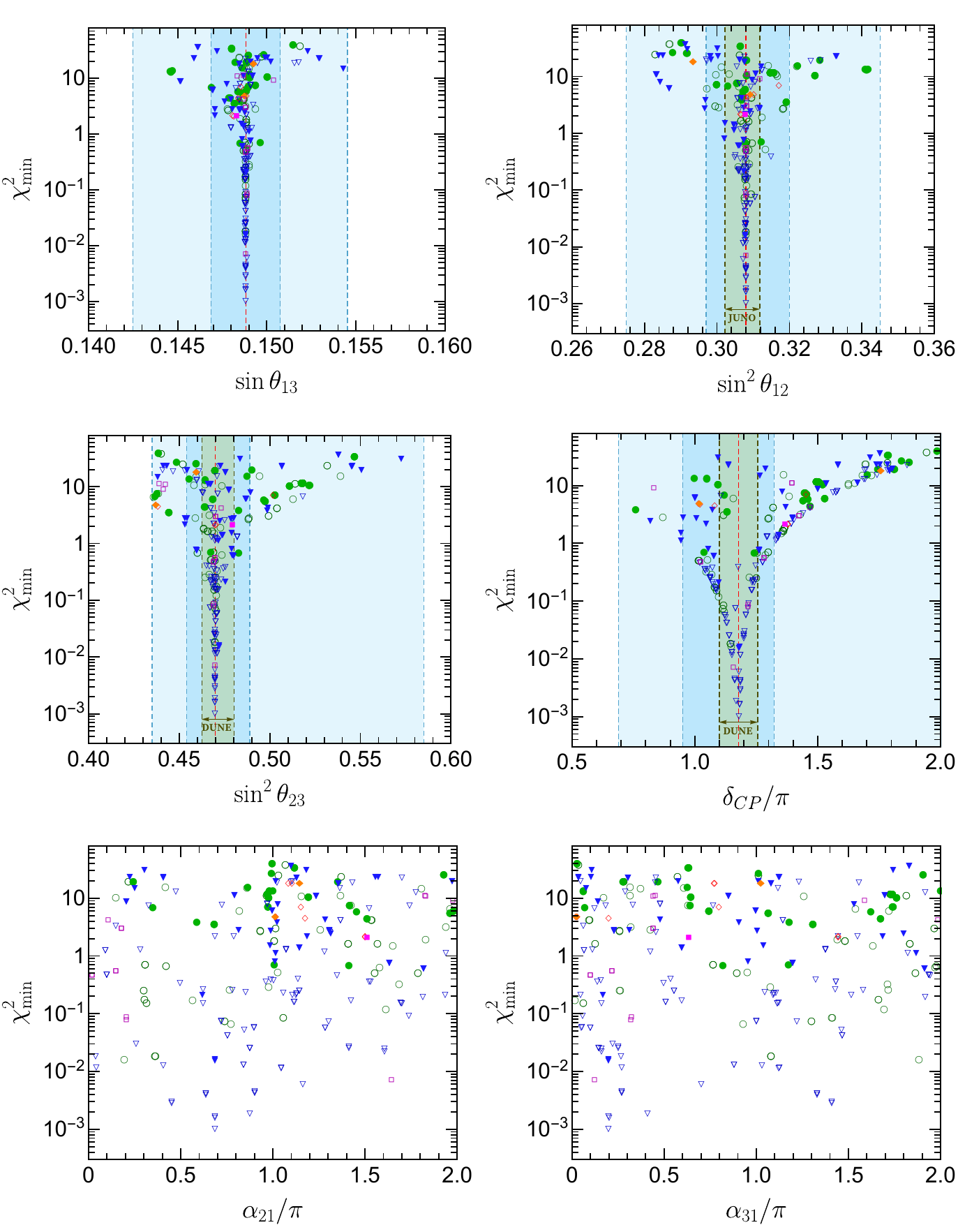}
\end{tabular}
\caption{\label{fig:bf_mixing}  The results of the best fit values of the  minimum value of $\chi^2$, the three lepton mixing angles and three CP violation phases for all 195 viable models. The red dashed lines in the first four panels represent the best fit values, and the light blue bounds represent the $1\sigma$ and $3\sigma$ ranges from NuFIT 6.0 with Super-Kamiokande atmospheric data~\cite{Esteban:2020cvm}. The lighter brown band in the panel of $\sin^{2}\theta_{12}$ is the prospective $3\sigma$ range after 6 years of JUNO running~\cite{JUNO:2022mxj}. The lighter green regions in the panels of $\sin^{2}\theta_{23}$ and $\delta_{CP}$ are the resolution in degrees after 15 years of DUNE running~\cite{DUNE:2020ypp} for true values of them corresponding to their  best fit values  given by NuFit 6.0. }
\end{figure}

\begin{figure}[t!]
\centering
\begin{tabular}{c}
\includegraphics[width=0.99\linewidth]{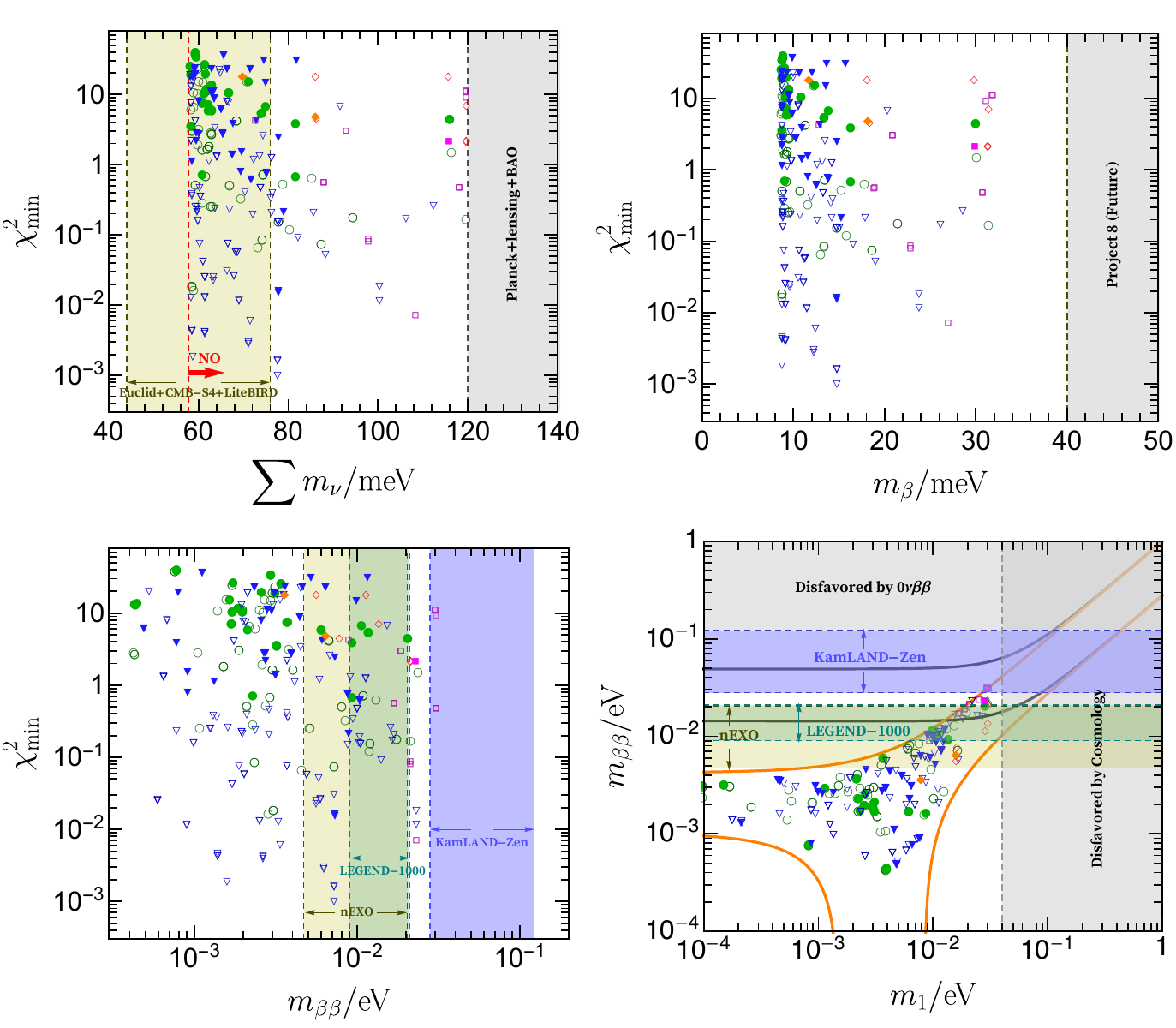}
\end{tabular}
\caption{\label{fig:bf_mass}  The best fit values of the  minimum value of $\chi^2$, the effective mass in $0\nu\beta\beta$-decay  $m_{\beta\beta}$, the kinematical mass in beta decay $m_{\beta}$ and the three neutrino mass sum $\sum m_{\nu}$. In the panel of the neutrino mass sum $\sum m_{\nu}$, the vertical bands indicate the current most stringent limit $\sum m_{\nu}<120\,\text{meV}$ from the Planck $+$ lensing $+$ BAO~\cite{Planck:2018vyg},  the next generation experiments sensitivity ranges $\sum m_{\nu}<(44-76)\,\text{meV}$ of Euclid+CMB-S4+LiteBIRD~\cite{Euclid:2024imf}, and the dashed line represents the limitation of the normal mass ordering ($\sum m_{\nu}\geq 57.75\,\text{meV}$). In the panel of the kinematical mass in beta decay $m_{\beta}$, the gray region represents Project 8 future bound ($m_{\beta}<0.04\,\text{meV}$)~\cite{Project8:2022wqh}. In the panel of the effective Majorana mass $m_{\beta\beta}$, the vertical bands indicate the latest result $m_{\beta\beta}<(28-122)\,\text{meV}$ of KamLAND-Zen~\cite{KamLAND-Zen:2024eml}, and the next generation experiments sensitivity ranges $m_{\beta\beta}<(9-21)\,\text{meV}$ from LEGEND-1000~\cite{LEGEND:2021bnm} and $m_{\beta\beta}<(4.7-20.3)\,\text{meV}$ from nEXO~\cite{nEXO:2021ujk}. }
\end{figure}

For each set values of the input parameters, we can obtain the predicted values of lepton masses, mixing parameters and the corresponding $\chi^2$. Then we can find out the lowest $\chi^2$ by extensively scanning the parameter space. Here it is sufficient to limit the modulus $\tau$ in the fundamental domain $\mathcal{D}=\left\{\tau\in\mathcal{H}\big||\Re(\tau)|\leq1/2, |\tau|\geq1\right\}$. When we scan over the input parameter space of the 80 Weinberg operator models and the 320 seesaw models without gCP symmetry, we find that 21 Weinberg operator models and 174 seesaw models which contain 76 models $C_{i}-D_{j}-N^{\prime}_{k}$  and 98 models $C^{\prime}_{i}-D^{\prime}_{j}-N_{k}$ can accommodate the experimental data, and all the fitting results for input parameters and predictions for  mixing angles,  CP violation phases, neutrino masses, the effective mass $m_{\beta\beta}$ in neutrinoless double beta decay ($0\nu\beta\beta$-decay) and the kinematical mass $m_{\beta}$  in beta decay of these models are summarized in tables~\ref{tab:WO_bf_noGCP}, \ref{tab:CiDjNpk_best_fit_noGCP} and \ref{tab:CpiDpjNk_best_fit_noGCP}, respectively. When the finite modular group $A_{5}$ is extended to combine with the gCP symmetry, only 4 of the 21 viable  Weinberg operator models, 48 of 76 seesaw models $C_{i}-D_{j}-N^{\prime}_{k}$  and 52 of 98 seesaw models $C^{\prime}_{i}-D^{\prime}_{j}-N_{k}$ can give results in agreement with the experimental data on mixing parameters and lepton masses. The best fit values of the input parameters and the observable quantities of them are listed in table~\ref{tab:WO_bf_GCP}, table~\ref{tab:CiDjNpk_best_fit_GCP} and table~\ref{tab:CpiDpjNk_best_fit_GCP}, respectively. Henceforth we focus on the predictions of models that give the three mixing angles $\theta_{13}$, $\theta_{12}$, $\theta_{23}$, the Dirac CP phase $\delta_{CP}$ and $\Delta m^{2}_{21}/m^{2}_{31}$ in their $3\sigma$ ranges of global data analysis~\cite{Esteban:2020cvm}, the mass ratios $m_{e}/m_{\mu}$, $m_{\mu}/m_{\tau}$ in the experimentally favored intervals shown in table~\ref{tab:bf_13sigma_data}, and the three neutrino mass sum below the current most stringent limit $\sum m_{\nu}<120\,\text{meV}$ from Planck $+$ lensing $+$ BAO~\cite{Planck:2018vyg}.

In all these models, the modulus $\tau$ is treated as an extra free parameter, varied to maximize the agreement between data and theoretical predictions. The best fit values of the complex $\tau$ in the fundamental domain of $SL(2,\mathbb{Z})$ for all the 195 viable models are displayed in figure~\ref{fig:bf_tau}. Furthermore, we can pin down the minimal value of $\chi^2$ for each model  relevant to three lepton mixing angles, three CP violation phases, the three neutrino mass sum $\sum m_{\nu}$, the effective mass in $0\nu\beta\beta$-decay  $m_{\beta\beta}$ and the kinematical mass in beta decay $m_{\beta}$. The results are shown in figure~\ref{fig:bf_mixing} and figure~\ref{fig:bf_mass}. From figure~\ref{fig:bf_mixing}, we find that the best fit values of the three mixing angles and Dirac CP phase of the most viable models  are predicted to lie in their experimentally allowed $1\sigma$ regions~\cite{Esteban:2020cvm}. These models can agree with the experimental data very well. In principle, if a very precise measurement of the lepton mixing parameters and neutrino masses is derived from the forthcoming  neutrino oscillation experiments and cosmic surveys, these data combined with a very accurate determination of $m_{\beta\beta}$ might provide a good opportunity for probing different modular models. The next generation medium baseline reactor neutrino experiment JUNO~\cite{JUNO:2022mxj} can make very precise measurement of the solar angle $\theta_{12}$. The prospective $3\sigma$ range of $\sin^2\theta_{12}$ after 6 years of JUNO running are displayed in figure~\ref{fig:bf_mixing}. Critical tests of the viability of these models will be provided by the planned high-precision measurements of $\theta_{23}$ and $\delta_{CP}$ at the future long baseline experiments DUNE~\cite{DUNE:2020ypp} and T2HK~\cite{Hyper-Kamiokande:2018ofw}. After 15 years of DUNE running, the resolution in degrees of them are plotted in figure~\ref{fig:bf_mixing}. We see that the synergy between  JUNO and long baseline neutrino experiments DUNE and T2HK will be extremely powerful for testing a large number of these modular models.

From figure~\ref{fig:bf_mass}, we find that the predictions for the neutrino mass sum $\sum m_{\nu}$ in most of the viable models can be tested by the next generation experiments sensitivity ranges $\sum m_{\nu}<(44-76)\,\text{meV}$ of Euclid+CMB-S4+LiteBIRD~\cite{Euclid:2024imf}. The best fit values of $m_{\beta}$ in all viable models are below the Project 8 future bound $0.04\text{eV}$~\cite{Project8:2022wqh}. The predictions for $m_{\beta\beta}$ in almost all viable models are compatible with the latest result $m_{\beta\beta}<(28-122)\,\text{meV}$ of KamLAND-Zen~\cite{KamLAND-Zen:2024eml}, and most of the viable models can be tested by the next generation experiments such as LEGEND-1000~\cite{LEGEND:2021bnm} which aims to improve the sensitivity to $m_{\beta\beta} < (9 \sim 21)\,$meV and  nEXO~\cite{nEXO:2021ujk} which expects to achieve $m_{\beta\beta} < (4.7 \sim 20.3)\,$meV. The relationship between the best fit values of the lightest neutrino mass $m_{1}$ and $m_{\beta\beta}$ in each viable model are also displayed in figure~\ref{fig:bf_mass}.

\section{\label{sec:example-models}Typical models}

\begin{figure}[t!]
\centering
\begin{tabular}{c}
\includegraphics[width=0.99\linewidth]{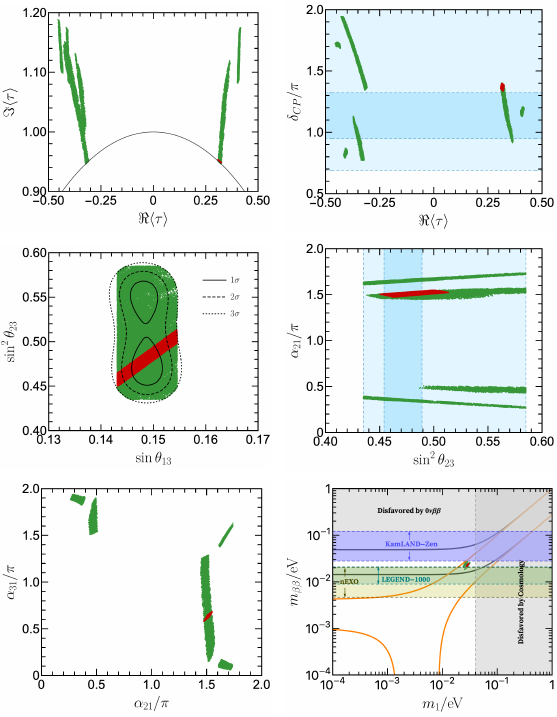}
\end{tabular}
\caption{\label{fig:WO_mixing}  The predicted correlations among the input free parameters, neutrino mixing angles and CP violation phases in the model $C^{\prime}_{2}-W^{\prime}_{4}$ with (red) and without (green) gCP symmetry. }
\end{figure}

From the discussion above, we see that the non-holomorphic $A_{5}$ modular models are quite predictive, particularly models with gCP symmetry. The reason is that the number of input parameter is less than the number of observable quantities. Hence the free parameters, mixing parameters and neutrino masses are generally correlated with each other.  In order to show the  predictions of non-holomorphic $A_{5}$ modular invariant models, we shall give detailed numerical results of Weinberg operator model $C^{\prime}_{2}-W^{\prime}_{4}$ and seesaw model $C^{\prime}_{6}-D^{\prime}_{1}-N_{1}$ as examples for illustration. The assignments of the two models are
\begin{equation}
L\sim \bm{3^{\prime}}, \qquad E^{c}_{1,2,3}\sim\bm{1}, \qquad k_{L}=1, \qquad k_{1}=-5, \qquad k_{2}=-3, \qquad k_{3}=1\,,
\end{equation}
and
\begin{equation}
N^{c}\sim\bm{3}, \qquad L\sim \bm{3^{\prime}}, \qquad E^{c}_{1,2,3}\sim\bm{1}, \qquad k_{N}=k_{L}=k_{1}=-2, \qquad k_{2}=4, \qquad k_{3}=6\,.
\end{equation}
For the two example models without gCP and with gCP, the best fit values of the free parameters, the neutrino masses and mixing parameters are  summarized in tables~\ref{tab:WO_bf_noGCP}, \ref{tab:WO_bf_GCP}, \ref{tab:CpiDpjNk_best_fit_noGCP} and \ref{tab:CpiDpjNk_best_fit_GCP}, respectively. After we comprehensively scan the parameter space of each model,  and require all the observables lie in their experimentally preferred $3\sigma$ regions, some interesting correlations among the input parameters and observables for the two example models are obtained. For model $C^{\prime}_{2}-W^{\prime}_{4}$, there are six independent and disconnected regions compatible with experiment data in the parameter space, and  the corresponding  correlations among input parameters and observables are plotted in figure~\ref{fig:WO_mixing}, where the results without gCP and with gCP are displayed in green and red, respectively. When gCP invariance is required in the model, all couplings are real and $\Re \tau$ is the unique source of CP violation. As a consequence, the three CP violation phases and the allowed regions of the input parameters are constrained to lie in narrow ranges in figure~\ref{fig:WO_mixing}. We find that the predicted ranges of the atmospheric mixing angle $\sin^2\theta_{23}$ and the Dirac CP violation phase $\delta_{CP}$ are $[0.4502,0.5134]$ and $[1.3367\pi,1.399\pi]$, respectively. These predictions may be tested at  future long-baseline experiments  DUNE~\cite{DUNE:2020ypp} and T2HK~\cite{Hyper-Kamiokande:2018ofw}, and at the discussed ESS$\nu$SB experiment~\cite{Alekou:2022emd}. The possible values of  the sum of neutrino masses $\sum m_{\nu}$ lie in the interval $[112\,\text{meV},120\,\text{meV}]$ which is below  the current most stringent limit $\sum m_{\nu}<120\,\text{meV}$  from the Planck $+$ lensing $+$ BAO collaboration~\cite{Planck:2018vyg}, and the allowed range of  the effective Majorana neutrino mass $m_{\beta\beta}$ is $[21.39\,\text{meV},24.79\,\text{meV}]$ which may be tested by the KamLAND-Zen experiment~\cite{KamLAND-Zen:2024eml}, and future large-scale $0\nu\beta\beta$-decay experiments, such as LEGEND-1000~\cite{LEGEND:2021bnm} and nEXO~\cite{nEXO:2021ujk}. Furthermore,  the two Majorana CP violation phases are predicted to be in narrow regions $\alpha_{21}\in[1.473\pi,1.549\pi]$ and $\alpha_{31}\in[0.5821\pi,0.6832\pi]$.

\begin{figure}[t!]
\centering
\begin{tabular}{c}
\includegraphics[width=0.99\linewidth]{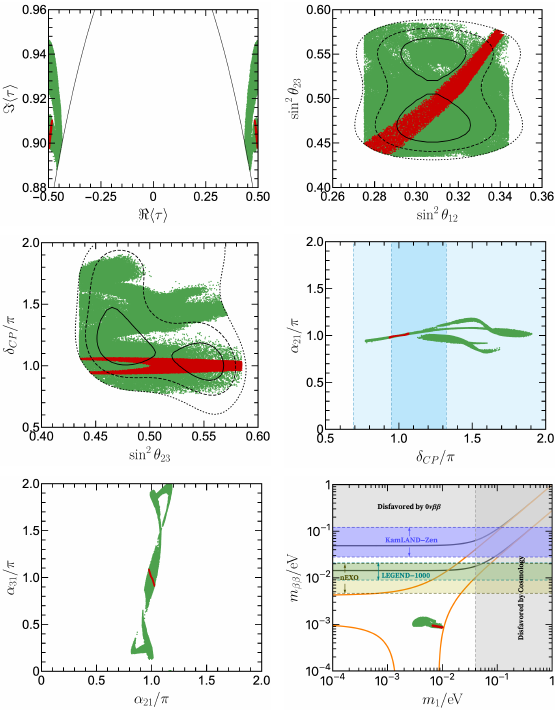}
\end{tabular}
\caption{\label{fig:seesaw_mixing}  The experimentally allowed values of the complex modulus $\tau$ and the correlations between  lepton mixing parameters in the model $C^{\prime}_{6}-D^{\prime}_{1}-N_{1}$ with (red) and without (green) gCP symmetry.}
\end{figure}

After we perform an extensive numerical scan over the parameter space of the seesaw model $C^{\prime}_{6}-D^{\prime}_{1}-N_{1}$, the numerical results are obtained. Then we find that the charged lepton mass hierarchies not require hierarchical values of the parameters $\alpha$, $\beta$ and $\gamma$. The allowed regions of the ratio $\beta/\alpha$ are $[1.189,2.899]$ and $[17.32,49.30]$, and the corresponding viable regions of the ratio $\gamma/\alpha$ are $[12.68,32.36]$ and $[0.636,1.938]$, respectively.  We  display allowed region of the modulus $\tau$ and the correlations  among different observables in figure~\ref{fig:seesaw_mixing}. From figure~\ref{fig:seesaw_mixing}, we find that  the atmospheric angle $\sin^2\theta_{23}$ are strongly correlated with the solar angle $\sin^2\theta_{12}$ when  gCP symmetry consistent with  non-holomorphic modular symmetry is imposed. It can be tested by JUNO experiment~\cite{JUNO:2022mxj} in combination with DUNE~\cite{DUNE:2020ypp} or T2HK~\cite{Hyper-Kamiokande:2018ofw}. Moreover, the three CP violation phases are all restricted to the narrow intervals $0.9339\pi\leq\delta_{CP}\leq1.066\pi$, $0.9749\pi\leq\alpha_{21}\leq1.026\pi$ and $0.9114\pi\leq\alpha_{31}\leq1.087\pi$. Hence model can be  tested at future long baseline neutrino oscillation experiments DUNE~\cite{DUNE:2020ypp} or T2HK~\cite{Hyper-Kamiokande:2018ofw}. The neutrino mass sum $\sum m_{\nu}$ and $m_{\beta\beta}$ are found to lie in narrow intervals $[66.87\,\text{meV},75.21\,\text{meV}]$ and $[0.8438\,\text{meV},0.9706\,\text{meV}]$, respectively. The prediction for the former is compatible with the upper bound on neutrino mass sum from Planck $+$ lensing $+$ BAO~\cite{Planck:2018vyg} and may be tested by Euclid+CMB-S4+LiteBIRD~\cite{Euclid:2024imf}, and the prediction for latter is much below the current most stringent limit from~\cite{KamLAND-Zen:2024eml}, and next generation $0\nu\beta\beta$-decay experiments LEGEND-1000~\cite{LEGEND:2021bnm} and nEXO~\cite{nEXO:2021ujk}.

\section{\label{sec:conclusion} Conclusion }

Modular symmetry is a promising approach for predicting both the hierarchical masses and flavor mixing parameters of fermions~\cite{Feruglio:2017spp}. In the present work, we have performed a comprehensive analysis of lepton models based on $\Gamma_{5}\cong A_{5}$ modular flavor symmetry in the framework of non-supersymmetric~\cite{Qu:2024rns}, and all the simplest lepton models without any other flavon except $\tau$ have been constructed.  In these models, the neutrinos are assumed to be Majorana particles. They are considered gaining masses via either the Weinberg operator or the type I seesaw mechanism. For all these models, the LH lepton doublets $L$ and the three RH charged leptons $E^{c}_{1,2,3}$ are assigned to a triplet  and singlet $\bm{1}$ of $A_{5}$ respectively, and the RH neutrinos $N^{c}$ transform as triplet $\bm{3}$ or $\bm{3^{\prime}}$ for seesaw models. The Yukawa couplings come from the polyharmonic Maa{\ss} forms of weights $k=\pm4,\,\pm2,\,0$ and level 5. Then  80 independent minimal Weinberg operator models and 320 independent minimal seesaw models are obtained, and the corresponding  charged lepton and neutrino masses matrices are shown in table~\ref{tab:sum_ch} and table~\ref{tab:sum_WO_SeeSaw}, respectively. All the 400 models depend on six real dimensionless free parameters (including the real and imaginary parts of $\tau$) in Eq.~\eqref{eq:six_inputs} and two overall parameters in the case of without gCP symmetry.  It is known that the gCP symmetry can be consistently combined with the non-holomorphic modular flavor symmetry~\cite{Qu:2024rns}. When the non-holomorphic modular flavor symmetry is extended to combine with the gCP symmetry, all coupling constants are enforced to be real in our working basis.

After performing an exhaustive numerical analysis for all the 400 models, we find out 21 simplest phenomenologically viable Weinberg operator models and 174 simplest phenomenologically viable seesaw models in the case of without gCP symmetry.  The best fit values of input parameters, lepton mixing angles, CP violation phases, neutrino masses, the $0\nu\beta\beta$-decay effective Majorana mass and the kinematical mass in beta decay are listed in tables~\ref{tab:WO_bf_noGCP},~\ref{tab:CiDjNpk_best_fit_noGCP} and~\ref{tab:CpiDpjNk_best_fit_noGCP}.   We find that the most viable models agree with the experimental data very well, in which all mixing parameters are predicted to lie in their experimentally allowed $1\sigma$ regions.  When gCP symmetry is imposed, then one more free parameter would be reduced  for all the 195 viable models. As a consequence, only 4 of the 21 viable Weinberg operator models and 100 of the 174 viable seesaw models can accommodate the experimental data in lepton sector, as shown in tables~\ref{tab:WO_bf_GCP},~\ref{tab:CiDjNpk_best_fit_GCP} and~\ref{tab:CpiDpjNk_best_fit_GCP}. The future medium baseline reactor neutrino oscillation experiments should measure  the solar mixing angle $\sin^{2}\theta_{12}$ with very good precision. A high precision determination of the atmospheric mixing $\sin^{2}\theta_{23}$ and the Dirac CP phase $\delta_{CP}$ can be performed by the next generation  long baseline neutrino experiments DUNE and T2HK. We find that the synergy between the forthcoming neutrino experiments JUNO and DUNE (or T2HK) will be extremely powerful for testing a large number of these viable modular models, particularly viable models with gCP symmetry. Additional important tests of the models will be provided by precision measurements of the neutrino mass sum $\sum m_{\nu}$ from the next generation experiments  Euclid+CMB-S4+LiteBIRD and the effective mass $m_{\beta\beta}$ from the future large-scale $0\nu\beta\beta$-decay experiments LEGEND-1000 and nEXO. Then the number of viable models will further be reduced.

Guided by the analysis of viable models,  we present detailed numerical results of Weinberg operator model $C^{\prime}_{2}-W^{\prime}_{4}$ and seesaw model $C^{\prime}_{6}-D^{\prime}_{1}-N_{1}$ as examples for illustration. The result predictions for lepton mixing parameters, neutrino masses and the $0\nu\beta\beta$-decay effective mass of the two models are obtained. When gCP symmetry is imposed, the allowed regions of the input parameters and mixing parameters are constrained to lie in narrow ranges, and some interesting correlations among the input parameters and observables for the two example models are obtained. These correlations could be tested by the next generation neutrino experiments.  The Weinberg operator model $C^{\prime}_{2}-W^{\prime}_{4}$ can be tested by the future large-scale $0\nu\beta\beta$-decay experiments LEGEND-1000 and nEXO.

\section*{Acknowledgements}

CCL is supported by Natural Science Basic Research Program of Shaanxi (Program No. 2024JC-YBQN-0004), and the National Natural Science Foundation of China under Grant No. 12247103. JNL is supported by the Grants No. NSFC-12147110 and the China Post-doctoral Science Foundation under Grant No. 2021M70. GJD is supported by the National Natural Science Foundation of China under Grant Nos.~12375104.

\newpage

\begin{center}
\renewcommand{\tabcolsep}{0.7mm}
\renewcommand{\arraystretch}{1.15}
\begin{small}
\setlength\LTcapwidth{\textwidth}
\setlength\LTleft{-0.0in}
\setlength\LTright{0pt}
\begin{longtable}{|c|c|c|c|c|c|c|c|c|c|c|c|c|c|c|c|c|c|c|c|c|c|c|c|c|c|c|c|c|c|c|c|c|c|c|c|c|c|c|c|c|c|c|c|c|c|c|c|c|c|c|c|c|c|c|c|c|c|c|c|c|c|c|c|c|c|c|c|c|c|c|c|c|c|c|c|c|c|c|c|c|c|c|c|c|c|c|c|c|c|c|c|c|c|c|c|c|c|c|c|c|c|c|c|c|c|c|c|c|c|c|c|c|c|c|c|c|c|}
\caption{\label{tab:WO_bf_noGCP}
The best fit values of the input parameters, three mixing angles $\theta_{12}$, $\theta_{13}$, $\theta_{23}$, Dirac CP violation phase $\delta_{CP}$, Majorana CP violation phases $\alpha_{21}$, $\alpha_{31}$,  three light neutrino masses $m_{1,2,3}$,  the effective mass $m_{\beta\beta}$ in $0\nu\beta\beta$-decay and the kinematical mass $m_{\beta}$ in beta decay  at the minimum values of $\chi^2$ for viable models $C_i-W_{j}$ and $C^{\prime}_i-W^{\prime}_{j}$ which are labeled as  $\{i,j\}$ in the case of without gCP symmetry. The best fit values of the charged lepton mass ratios $m_{e}/m_{\mu}$ and  $m_{\mu}/m_{\tau}$ are both the global best fit values of them, i.e. $m_{e}/m_{\mu}=0.004737$ and $m_{\mu}/m_{\tau}=0.05882$. The best fit values of the mass sum $\sum m_{\nu}$ and $\Delta m^{2}_{21}/\Delta m^{2}_{31}$ can be easy obtained from the three neutrino masses. In the following tables, we would not show them as the same reason.} \\
\midrule
\specialrule{0em}{1.0pt}{1.0pt}

\endfirsthead

\multicolumn{100}{c}
{{\bfseries \tablename\ \thetable{} -- continued from previous page}} \\
\hline

\endhead

\caption[]{continues on next page}\\
\endfoot

\endlastfoot

\hline
\texttt{Models} & \multicolumn{99}{c|}{ Best fit results for 8 viable models $C_i-W_{j}$ without gCP}   \\ \hline
$ \{i,j\}$      &  \multicolumn{11}{c|}{  $\Re\langle \tau \rangle$} &  \multicolumn{11}{c|}{  $\Im\langle \tau \rangle$}  &  \multicolumn{11}{c|}{  $\beta/\alpha$} &  \multicolumn{11}{c|}{  $\gamma/\alpha$} &  \multicolumn{11}{c|}{  $|g_{2}/g_{1}|$} &  \multicolumn{11}{c|}{  $\text{arg}(g_{2}/g_{1})/\pi$} &  \multicolumn{11}{c|}{  $\alpha v$/MeV} & \multicolumn{11}{c|}{  $(g_{1}v^2/\Lambda)$/meV} &  \multicolumn{11}{c|}{$\chi^2_{\mathrm{min}}$} \\ \hline

 \{1,1\} & \multicolumn{11}{c|}{ $-$0.115 } & \multicolumn{11}{c|}{ 1.18 } & \multicolumn{11}{c|}{ 489.1 } & \multicolumn{11}{c|}{ 8.29 } & \multicolumn{11}{c|}{ 0.441 } & \multicolumn{11}{c|}{ 0.296 } & \multicolumn{11}{c|}{ 6.74 } & \multicolumn{11}{c|}{ 67.4 } & \multicolumn{11}{c|}{ 2.20 } \\ \hline
 \{1,3\} & \multicolumn{11}{c|}{ $-$0.190 } & \multicolumn{11}{c|}{ 1.27 } & \multicolumn{11}{c|}{ 11.4 } & \multicolumn{11}{c|}{ 0.00448 } & \multicolumn{11}{c|}{ 0.725 } & \multicolumn{11}{c|}{ 0.170 } & \multicolumn{11}{c|}{ 242.8 } & \multicolumn{11}{c|}{ 27.1 } & \multicolumn{11}{c|}{ 18.3 } \\ \hline
 \{2,1\} & \multicolumn{11}{c|}{ $-$0.348 } & \multicolumn{11}{c|}{ 1.11 } & \multicolumn{11}{c|}{ 10.3 } & \multicolumn{11}{c|}{ 0.000907 } & \multicolumn{11}{c|}{ 0.246 } & \multicolumn{11}{c|}{ 0.151 } & \multicolumn{11}{c|}{ 362.9 } & \multicolumn{11}{c|}{ 116.8 } & \multicolumn{11}{c|}{ 7.11 } \\ \hline
 \{2,3\} & \multicolumn{11}{c|}{ $-$0.190 } & \multicolumn{11}{c|}{ 1.27 } & \multicolumn{11}{c|}{ 11.4 } & \multicolumn{11}{c|}{ 0.00149 } & \multicolumn{11}{c|}{ 0.545 } & \multicolumn{11}{c|}{ 0.276 } & \multicolumn{11}{c|}{ 242.8 } & \multicolumn{11}{c|}{ 36.0 } & \multicolumn{11}{c|}{ 18.3 } \\
\end{longtable}
\end{small}
\end{center}

\begin{table}[h!]
        \centering
\small
                \renewcommand{\tabcolsep}{0.7mm}
        \renewcommand{\arraystretch}{1.1}
\caption{\label{tab:WO_bf_GCP}The best fit values of the input parameters, lepton masses and mixing parameters at the minimum values of $\chi^2$ for viable models $C_i-W_{j}$ and $C^{\prime}_i-W^{\prime}_{j}$  with gCP symmetry.}
\begin{tabular}{|c|c|c|c|c|c|c|c|c|c|c|c|c|c|c|c|c|c|c|c|c|c|c|c|c|c|c|c|c|c|c|c|c|c|c|c|c|c|c|c|c|c|c|c|c|c|c|c|c|c|c|c|c|c|c|c|c|c|c|c|c|c|c|c|c|c|c|c|c|c|c|c|c|c|c|c|c|c|c|c|c|c|c|c|c|c|c|c|c|c|c|c|c|c|c|c|c|c|c|c|c|c|c|c|c|c|c|c|c|c|c|c|c|c|c|c|c|c|}  \hline\hline
\texttt{Models} &  \multicolumn{88}{c|}{ Best fit results for 3 viable models $C_i-W_{j}$ with gCP}  \\ \hline
$ \{i,j\}$ & \multicolumn{11}{c|}{$\Re\langle \tau \rangle$} & \multicolumn{11}{c|}{$\Im\langle \tau \rangle$}  & \multicolumn{11}{c|}{$\beta/\alpha$} & \multicolumn{11}{c|}{$\gamma/\alpha$} & \multicolumn{11}{c|}{$g_{2}/g_{1}$} & \multicolumn{11}{c|}{$\alpha v$/MeV} & \multicolumn{11}{c|}{$(g_{1}v^2/\Lambda)/$meV} &  \multicolumn{11}{c|}{$\chi^2_{\mathrm{min}}$}\\ \hline
  \{1,3\} & \multicolumn{11}{c|}{ $-$0.190 } & \multicolumn{11}{c|}{ 1.27 } & \multicolumn{11}{c|}{ 11.4 } & \multicolumn{11}{c|}{ 0.00448 } & \multicolumn{11}{c|}{ 0.842 } & \multicolumn{11}{c|}{ 242.8 } & \multicolumn{11}{c|}{ 23.3 } & \multicolumn{11}{c|}{ 18.3 } \\ \hline
 \{2,3\} & \multicolumn{11}{c|}{ $-$0.190 } & \multicolumn{11}{c|}{ 1.27 } & \multicolumn{11}{c|}{ 11.4 } & \multicolumn{11}{c|}{ 0.00149 } & \multicolumn{11}{c|}{ 0.842 } & \multicolumn{11}{c|}{ 242.8 } & \multicolumn{11}{c|}{ 23.3 } & \multicolumn{11}{c|}{ 18.3 } \\ \hline
 \{3,1\} & \multicolumn{11}{c|}{ $-$0.499 } & \multicolumn{11}{c|}{ 1.01 } & \multicolumn{11}{c|}{ 0.00511 } & \multicolumn{11}{c|}{ 3.28 } & \multicolumn{11}{c|}{ 0.468 } & \multicolumn{11}{c|}{ 579.9 } & \multicolumn{11}{c|}{ 138.1 } & \multicolumn{11}{c|}{ 4.90 } \\
\hline \hline
\end{tabular}
\end{table}

\newpage

\begin{center}
\renewcommand{\tabcolsep}{0.7mm}
\renewcommand{\arraystretch}{1.15}
\begin{small}
\setlength\LTcapwidth{\textwidth}
\setlength\LTleft{-0.1in}
\setlength\LTright{0pt}
\begin{longtable}{|c|c|c|c|c|c|c|c|c|c|c|c|c|c|c|c|c|c|c|c|c|c|c|c|c|c|c|c|c|c|c|c|c|c|c|c|c|c|c|c|c|c|c|c|c|c|c|c|c|c|c|c|c|c|c|c|c|c|c|c|c|c|c|c|c|c|c|c|c|c|c|c|c|c|c|c|c|c|c|c|c|c|c|c|c|c|c|c|c|c|c|c|c|c|c|c|c|c|c|c|c|c|c|c|c|c|c|c|c|c|c|c|c|c|c|c|c|c|}
\caption{\label{tab:CiDjNpk_best_fit_noGCP}
The best fit values of the input parameters, neutrino masses and mixing parameters at the minimum values of $\chi^2$ for viable models $C_i-D_{j}-N^{\prime}_{k}$ which are labeled as $\{i,j,k\}$. In this table, the gCP symmetry compatible with $A_{5}$ is not imposed.} \\
\midrule
\specialrule{0em}{1.0pt}{1.0pt}

\endfirsthead

\multicolumn{100}{c}
{{\bfseries \tablename\ \thetable{} -- continued from previous page}} \\
\hline

\endhead

\caption[]{continues on next page}\\
\endfoot

\endlastfoot

\hline
\texttt{Models} & \multicolumn{99}{c|}{Best fit results for 76 viable models $C_{i}-D_{j}-N^{\prime}_{k}$ without  gCP}  \\ \hline
 $\{i,j,k\}$ &  \multicolumn{11}{c|}{ $\Re\langle \tau \rangle$} & \multicolumn{11}{c|}{ $\Im\langle \tau \rangle$}  &  \multicolumn{11}{c|}{ $\beta/\alpha$} &  \multicolumn{11}{c|}{ $\gamma/\alpha$} &  \multicolumn{11}{c|}{ $|h_{2}/h_{1}|$} &  \multicolumn{11}{c|}{ $\text{arg}(h_{2}/h_{1})/\pi$} &  \multicolumn{11}{c|}{ $\alpha v$/MeV} &  \multicolumn{11}{c|}{ $(g_{1}^2v^2/(h_{1}\Lambda))/$meV} & \multicolumn{11}{c|}{$\chi^2_{\mathrm{min}}$ } \\ \hline

\{1,1,1\} & \multicolumn{11}{c|}{ 0.289 } & \multicolumn{11}{c|}{ 1.28 } & \multicolumn{11}{c|}{ 829.6 } & \multicolumn{11}{c|}{ 19.7 } & \multicolumn{11}{c|}{ 1.01 } & \multicolumn{11}{c|}{ 1.97 } & \multicolumn{11}{c|}{ 3.27 } & \multicolumn{11}{c|}{ 47.0 } & \multicolumn{11}{c|}{ 10.3 } \\ \hline
 \{1,1,2\} & \multicolumn{11}{c|}{ 0.465 } & \multicolumn{11}{c|}{ 1.09 } & \multicolumn{11}{c|}{ 1237. } & \multicolumn{11}{c|}{ 17.2 } & \multicolumn{11}{c|}{ 2.54 } & \multicolumn{11}{c|}{ 0.640 } & \multicolumn{11}{c|}{ 3.15 } & \multicolumn{11}{c|}{ 428.0 } & \multicolumn{11}{c|}{ 0.288 } \\ \hline
 \{1,1,3\} & \multicolumn{11}{c|}{ $-$0.0397 } & \multicolumn{11}{c|}{ 1.17 } & \multicolumn{11}{c|}{ 323.0 } & \multicolumn{11}{c|}{ 5.23 } & \multicolumn{11}{c|}{ 0.949 } & \multicolumn{11}{c|}{ 0.0132 } & \multicolumn{11}{c|}{ 10.5 } & \multicolumn{11}{c|}{ 159.6 } & \multicolumn{11}{c|}{ 0.508 } \\ \hline
 \{1,1,4\} & \multicolumn{11}{c|}{ $-$0.326 } & \multicolumn{11}{c|}{ 1.11 } & \multicolumn{11}{c|}{ 10.6 } & \multicolumn{11}{c|}{ 0.00156 } & \multicolumn{11}{c|}{ 0.897 } & \multicolumn{11}{c|}{ 0.308 } & \multicolumn{11}{c|}{ 352.7 } & \multicolumn{11}{c|}{ 331.1 } & \multicolumn{11}{c|}{ 0.680 } \\
\end{longtable}
\end{small}
\end{center}

\newpage

\begin{center}
\renewcommand{\tabcolsep}{0.7mm}
\renewcommand{\arraystretch}{1.15}
\begin{small}
\setlength\LTcapwidth{\textwidth}
\setlength\LTleft{0.0in}
\setlength\LTright{0pt}
\begin{longtable}{|c|c|c|c|c|c|c|c|c|c|c|c|c|c|c|c|c|c|c|c|c|c|c|c|c|c|c|c|c|c|c|c|c|c|c|c|c|c|c|c|c|c|c|c|c|c|c|c|c|c|c|c|c|c|c|c|c|c|c|c|c|c|c|c|c|c|c|c|c|c|c|c|c|c|c|c|c|c|c|c|c|c|c|c|c|c|c|c|c|c|c|c|c|c|c|c|c|c|c|c|c|c|c|c|c|c|c|c|c|c|c|c|c|c|c|c|c|c|c|c|c|c|c|c|c|c|c|c|c|c|c|c|c|c|c|c|c|c|c|c|c|c|c|c|c|c|c|c|c|c|c|c|c|c|c|c|c|c|c|c|c|c|c|}
\caption{\label{tab:CiDjNpk_best_fit_GCP}
The best fit values of the input parameters, neutrino masses and mixing parameters at the minimum values of $\chi^2$ for viable models $C_i-D_{j}-N^{\prime}_{k}$ with gCP symmetry. } \\
\midrule
\specialrule{0em}{1.0pt}{1.0pt}
                        
\endfirsthead
                        
\multicolumn{89}{c}
{{\bfseries \tablename\ \thetable{} -- continued from previous page}} \\
        \hline

\endhead
                        
                \caption[]{continues on next page}\\
                        \endfoot
                        
        \endlastfoot
                        
\hline
\texttt{Models} & \multicolumn{88}{c|}{Best fit results for 48 viable  models $C_{i}-D_{j}-N^{\prime}_{k}$ with gCP}   \\ \hline
$\{i,j,k\}$ & \multicolumn{11}{c|}{$\Re\langle \tau \rangle$} & \multicolumn{11}{c|}{$\Im\langle \tau \rangle$}  & \multicolumn{11}{c|}{$\beta/\alpha$} & \multicolumn{11}{c|}{$\gamma/\alpha$} & \multicolumn{11}{c|}{$h_{2}/h_{1}$}  &  \multicolumn{11}{c|}{ $\alpha v$/MeV} & \multicolumn{11}{c|}{$(g_{1}^2v^2/(h_{1}\Lambda))/$meV} &\multicolumn{11}{c|}{$\chi^2_{\mathrm{min}}$} \\ \hline

  \{1,1,2\} & \multicolumn{11}{c|}{ 0.435 } & \multicolumn{11}{c|}{ 1.08 } & \multicolumn{11}{c|}{ 1219. } & \multicolumn{11}{c|}{ 16.8 } & \multicolumn{11}{c|}{ $-$5.18 } & \multicolumn{11}{c|}{ 3.20 } & \multicolumn{11}{c|}{ 877.7 } & \multicolumn{11}{c|}{ 0.714 } \\ \hline
 \{1,1,3\} & \multicolumn{11}{c|}{ 0.108 } & \multicolumn{11}{c|}{ 1.36 } & \multicolumn{11}{c|}{ 537.5 } & \multicolumn{11}{c|}{ 16.9 } & \multicolumn{11}{c|}{ 2.83 } & \multicolumn{11}{c|}{ 4.23 } & \multicolumn{11}{c|}{ 49.3 } & \multicolumn{11}{c|}{ 25.8 } \\ \hline
 \{1,3,1\} & \multicolumn{11}{c|}{ 0.0244 } & \multicolumn{11}{c|}{ 1.24 } & \multicolumn{11}{c|}{ 390.3 } & \multicolumn{11}{c|}{ 7.96 } & \multicolumn{11}{c|}{ 0.808 } & \multicolumn{11}{c|}{ 7.51 } & \multicolumn{11}{c|}{ 67.2 } & \multicolumn{11}{c|}{ 4.49 } \\ \hline
 \{1,3,2\} & \multicolumn{11}{c|}{ $-$0.0730 } & \multicolumn{11}{c|}{ 1.17 } & \multicolumn{11}{c|}{ 0.00123 } & \multicolumn{11}{c|}{ 0.586 } & \multicolumn{11}{c|}{ 1.77 } & \multicolumn{11}{c|}{ 1315. } & \multicolumn{11}{c|}{ 111.5 } & \multicolumn{11}{c|}{ 3.88 } \\ \hline
 \{2,2,2\} & \multicolumn{11}{c|}{ $-$0.462 } & \multicolumn{11}{c|}{ 1.03 } & \multicolumn{11}{c|}{ 0.00173 } & \multicolumn{11}{c|}{ 1.09 } & \multicolumn{11}{c|}{ 1.45 } & \multicolumn{11}{c|}{ 758.8 } & \multicolumn{11}{c|}{ 235.5 } & \multicolumn{11}{c|}{ 15.6 } \\
\end{longtable}
        \end{small}
\end{center}

\newpage

\begin{center}
\renewcommand{\tabcolsep}{0.7mm}
\renewcommand{\arraystretch}{1.15}
\begin{small}
\setlength\LTcapwidth{\textwidth}
\setlength\LTleft{-0.1in}
\setlength\LTright{0pt}
\begin{longtable}{|c|c|c|c|c|c|c|c|c|c|c|c|c|c|c|c|c|c|c|c|c|c|c|c|c|c|c|c|c|c|c|c|c|c|c|c|c|c|c|c|c|c|c|c|c|c|c|c|c|c|c|c|c|c|c|c|c|c|c|c|c|c|c|c|c|c|c|c|c|c|c|c|c|c|c|c|c|c|c|c|c|c|c|c|c|c|c|c|c|c|c|c|c|c|c|c|c|c|c|c|c|c|c|c|c|c|c|c|c|c|c|c|c|c|c|c|c|c|}
\caption{\label{tab:CpiDpjNk_best_fit_noGCP}
The best fit values of the input parameters, neutrino masses and mixing parameters at the minimum values of $\chi^2$ for viable models $C^{\prime}_i-D^{\prime}_{j}-N_{k}$ without gCP symmetry. } \\
\midrule
\specialrule{0em}{1.0pt}{1.0pt}
                        
\endfirsthead
                        
\multicolumn{99}{c}
{{\bfseries \tablename\ \thetable{} -- continued from previous page}} \\
\hline
                        
\endhead
                        
\caption[]{continues on next page}\\
\endfoot
\endlastfoot
\hline
\texttt{Models} &  \multicolumn{99}{c|}{ Best fit results for 98 viable models $C^{\prime}_i-D^{\prime}_{j}-N_{k}$  without gCP}  \\ \hline
$\{i,j,k\}$     & \multicolumn{11}{c|}{$\Re\langle \tau \rangle$} & \multicolumn{11}{c|}{$\Im\langle \tau \rangle$}  & \multicolumn{11}{c|}{$\beta/\alpha$} & \multicolumn{11}{c|}{$\gamma/\alpha$} & \multicolumn{11}{c|}{$|h_{2}/h_{1}|$} & \multicolumn{11}{c|}{$\text{arg}(h_{2}/h_{1})/\pi$} &  \multicolumn{11}{c|}{ $\alpha v$/MeV} & \multicolumn{11}{c|}{$(g_{1}^2v^2/(h_{1}\Lambda))/$meV} & \multicolumn{11}{c|}{$\chi^2_{\mathrm{min}}$}\\ \hline

 \{1,1,1\} & \multicolumn{11}{c|}{ $-$0.232 } & \multicolumn{11}{c|}{ 1.08 } & \multicolumn{11}{c|}{ 40.3 } & \multicolumn{11}{c|}{ 1215. } & \multicolumn{11}{c|}{ 0.763 } & \multicolumn{11}{c|}{ 2.00 } & \multicolumn{11}{c|}{ 5.67 } & \multicolumn{11}{c|}{ 146.7 } & \multicolumn{11}{c|}{ 0.0920 } \\ \hline
 \{1,1,2\} & \multicolumn{11}{c|}{ 0.422 } & \multicolumn{11}{c|}{ 1.03 } & \multicolumn{11}{c|}{ 8.96 } & \multicolumn{11}{c|}{ 0.0122 } & \multicolumn{11}{c|}{ 0.317 } & \multicolumn{11}{c|}{ 1.97 } & \multicolumn{11}{c|}{ 389.5 } & \multicolumn{11}{c|}{ 98.4 } & \multicolumn{11}{c|}{ 2.60 } \\ \hline
 \{1,1,3\} & \multicolumn{11}{c|}{ 0.354 } & \multicolumn{11}{c|}{ 1.06 } & \multicolumn{11}{c|}{ 79.7 } & \multicolumn{11}{c|}{ 2381. } & \multicolumn{11}{c|}{ 1.62 } & \multicolumn{11}{c|}{ 1.98 } & \multicolumn{11}{c|}{ 3.30 } & \multicolumn{11}{c|}{ 161.8 } & \multicolumn{11}{c|}{ 7.76 } \\ \hline
 \{1,1,4\} & \multicolumn{11}{c|}{ 0.183 } & \multicolumn{11}{c|}{ 1.06 } & \multicolumn{11}{c|}{ 28.1 } & \multicolumn{11}{c|}{ 930.9 } & \multicolumn{11}{c|}{ 0.220 } & \multicolumn{11}{c|}{ 0.310 } & \multicolumn{11}{c|}{ 7.54 } & \multicolumn{11}{c|}{ 55.5 } & \multicolumn{11}{c|}{ 0.470 } \\
\end{longtable}
        \end{small}
\end{center}

\newpage

\begin{center}
\renewcommand{\tabcolsep}{0.7mm}
\renewcommand{\arraystretch}{1.15}
\begin{small}
\setlength\LTcapwidth{\textwidth}
\setlength\LTleft{0.0in}
\setlength\LTright{0pt}
\begin{longtable}{|c|c|c|c|c|c|c|c|c|c|c|c|c|c|c|c|c|c|c|c|c|c|c|c|c|c|c|c|c|c|c|c|c|c|c|c|c|c|c|c|c|c|c|c|c|c|c|c|c|c|c|c|c|c|c|c|c|c|c|c|c|c|c|c|c|c|c|c|c|c|c|c|c|c|c|c|c|c|c|c|c|c|c|c|c|c|c|c|c|c|c|c|c|c|c|c|c|c|c|c|c|c|c|c|c|c|c|c|c|c|c|c|c|c|c|c|c|c|c|c|c|c|c|c|c|c|c|c|c|c|c|c|c|c|c|c|c|c|c|c|c|c|c|c|c|c|c|c|c|c|c|c|c|c|c|c|c|c|c|c|c|c|c|}
\caption{\label{tab:CpiDpjNk_best_fit_GCP} The best fit values of the input parameters, neutrino masses and mixing parameters at the minimum values of $\chi^2$ for viable models $C^{\prime}_i-D^{\prime}_{j}-N_{k}$ with gCP symmetry. } \\
\midrule
\specialrule{0em}{1.0pt}{1.0pt}
                        
\endfirsthead
                        
\multicolumn{89}{c}{{\bfseries \tablename\ \thetable{} -- continued from previous page}} \\ \hline

\endhead
                        
\caption[]{continues on next page}\\
\endfoot
                        
\endlastfoot \hline
\texttt{Models} & \multicolumn{88}{c|}{ Best fit results for 52 models $C^{\prime}_i-D^{\prime}_{j}-N_{k}$ with gCP}   \\ \hline
$\{i,j,k\}$& \multicolumn{11}{c|}{$\Re\langle \tau \rangle$} & \multicolumn{11}{c|}{$\Im\langle \tau \rangle$}  & \multicolumn{11}{c|}{$\beta/\alpha$} & \multicolumn{11}{c|}{$\gamma/\alpha$} & \multicolumn{11}{c|}{$h_{2}/h_{1}$}  &  \multicolumn{11}{c|}{ $\alpha v$/MeV}  & \multicolumn{11}{c|}{$(g_{1}^2v^2/(h_{1}\Lambda))/$meV} & \multicolumn{11}{c|}{$\chi^2_{\mathrm{min}}$} \\ \hline

 \{1,1,2\} & \multicolumn{11}{c|}{ 0.465 } & \multicolumn{11}{c|}{ 1.03 } & \multicolumn{11}{c|}{ 9.04 } & \multicolumn{11}{c|}{ 0.0117 } & \multicolumn{11}{c|}{ 0.289 } & \multicolumn{11}{c|}{ 392.0 } & \multicolumn{11}{c|}{ 90.6 } & \multicolumn{11}{c|}{ 4.30 } \\ \hline
 \{1,1,4\} & \multicolumn{11}{c|}{ 0.402 } & \multicolumn{11}{c|}{ 0.986 } & \multicolumn{11}{c|}{ 0.00273 } & \multicolumn{11}{c|}{ 20.0 } & \multicolumn{11}{c|}{ $-$0.00824 } & \multicolumn{11}{c|}{ 509.2 } & \multicolumn{11}{c|}{ 22.4 } & \multicolumn{11}{c|}{ 8.04 } \\ \hline
 \{1,3,1\} & \multicolumn{11}{c|}{ $-$0.185 } & \multicolumn{11}{c|}{ 1.21 } & \multicolumn{11}{c|}{ 38.3 } & \multicolumn{11}{c|}{ 703.0 } & \multicolumn{11}{c|}{ 1.02 } & \multicolumn{11}{c|}{ 7.10 } & \multicolumn{11}{c|}{ 40.7 } & \multicolumn{11}{c|}{ 0.620 } \\ \hline
 \{1,3,3\} & \multicolumn{11}{c|}{ $-$0.277 } & \multicolumn{11}{c|}{ 1.16 } & \multicolumn{11}{c|}{ 0.0585 } & \multicolumn{11}{c|}{ 0.00187 } & \multicolumn{11}{c|}{ 1.63 } & \multicolumn{11}{c|}{ 4024. } & \multicolumn{11}{c|}{ 15.6 } & \multicolumn{11}{c|}{ 2.20 } \\ \hline
 \{2,1,1\} & \multicolumn{11}{c|}{ 0.476 } & \multicolumn{11}{c|}{ 0.910 } & \multicolumn{11}{c|}{ 4.88 } & \multicolumn{11}{c|}{ 28.2 } & \multicolumn{11}{c|}{ 0.194 } & \multicolumn{11}{c|}{ 51.7 } & \multicolumn{11}{c|}{ 42.7 } & \multicolumn{11}{c|}{ 3.89 } \\
\end{longtable}
        \end{small}
\end{center}

\newpage

\section*{Appendix}

\begin{appendix}

\section{\label{sec:A5_group_theory}Finite modular group $\Gamma_5\cong A_{5}$}

The finite modular group $\Gamma_5\cong A_{5}$ can be generated by the modular generators $S$ and $T$ which satisfy the multiplication rules:
\begin{equation}
        S^2=T^5=(ST)^3=1\,.
\end{equation}
The $A_5$ group has five irreducible representations: one singlet representation $\bm{1}$, two three-dimensional representations $\bm{3}$ and $\bm{3^{\prime}}$, one four-dimensional representation $\bm{4}$ and one five-dimensional representation $\bm{5}$. In the present work, we shall follow the conventions of Refs.~\cite{Ding:2011cm,Li:2015jxa,Novichkov:2018nkm,Ding:2019xna} and  the explicit forms of the generators $S$ and $T$ in the five irreducible representations are as follows:
\begin{eqnarray}
        \begin{array}{cll}
                \bm{1:} &   ~S=1\,,~ &  T=1 \,, ~\\[-14pt] \\[4pt]
                \bm{3:} &   ~S=\frac{1}{\sqrt{5}}
                \begin{pmatrix}
                        1 & -\sqrt{2} & -\sqrt{2} \\
                        -\sqrt{2} & -\phi  & \phi-1 \\
                        -\sqrt{2} & \phi-1 & -\phi
                \end{pmatrix}\,,~
                & T=\text{diag}(1,\omega_{5},\omega^4_{5})
                \,,~~\\[-14pt] \\[4pt]
                \bm{3^{\prime}:} &  ~S=\frac{1}{\sqrt{5}}
                \begin{pmatrix}
                        -1 & \sqrt{2} & \sqrt{2} \\
                        \sqrt{2} & 1-\phi & \phi  \\
                        \sqrt{2} & \phi  & 1-\phi
                \end{pmatrix}\,, ~
                &T=
        \text{diag}(1,\omega^2_{5},\omega^3_{5})
                \,,~~\\[-14pt] \\[4pt]
                \bm{4:} &  ~S=\frac{1}{\sqrt{5}}
                \begin{pmatrix}
                        1 & \phi-1 & \phi  & -1 \\
                        \phi-1 & -1 & 1 & \phi  \\
                        \phi  & 1 & -1 & \phi-1 \\
                        -1 & \phi  & \phi-1 & 1
                \end{pmatrix}\,,~
                &T=
        \text{diag}(\omega_{5},\omega^2_{5},\omega^3_{5},\omega^4_{5})\,,~~\\[-14pt] \\[4pt]
                \bm{5:} &  ~S=\frac{1}{5}
                \begin{pmatrix}
                        -1 & \sqrt{6} & \sqrt{6} & \sqrt{6} & \sqrt{6} \\
                        \sqrt{6} & (\phi-1)^{2} & -2 \phi  & 2(\phi-1) & \phi ^2 \\
                        \sqrt{6} & -2\phi  & \phi^2& (\phi-1)^{2} & 2(\phi-1) \\
                        \sqrt{6} & 2(\phi-1) & (\phi-1)^{2} & \phi ^2 & -2 \phi  \\
                        \sqrt{6} & \phi^2 & 2(\phi-1) & -2 \phi  & (\phi-1)^{2}
                \end{pmatrix}\,,~
                &T=
                \text{diag}(1,\omega_{5},\omega^2_{5},\omega^3_{5},\omega^4_{5})\,,~~
        \end{array}
\end{eqnarray}
where $\omega_5=e^{2\pi i/5}$ denotes the quintic root of unit and $\phi=(1+\sqrt{5})/2$ is the golden ratio. Then the Kronecker products between various irreducible representations follow immediately
\begin{eqnarray}
\nonumber&&\bm{1}\otimes \bm{R}=\bm{R}\otimes\bm{1}=\bm{R},~~~\bm{3}\otimes\bm{3}=\bm{1_{S}}\oplus\bm{3_{A}}\oplus\bm{5_{S}},~~~\bm{3^{\prime}}\otimes\bm{3^{\prime}}=\bm{1_{S}}\oplus\bm{3^{\prime}_{A}}\oplus\bm{5_{S}}, ~~~\bm{3}\times\bm{3^{\prime}}=\bm{4}\oplus\bm{5},\\
        \nonumber&&\bm{3}\otimes\bm{4}=\bm{3^{\prime}}\oplus\bm{4}\oplus\bm{5},~~~\bm{3^{\prime}}\otimes\bm{4}=\bm{3}\oplus\bm{4}\oplus\bm{5},~~~\bm{3}\otimes\bm{5}
        =\bm{3}\oplus\bm{3^{\prime}}\oplus\bm{4}\oplus\bm{5},\\
        \nonumber&&\bm{3^{\prime}}\otimes\bm{5}=\bm{3}\oplus\bm{3^{\prime}}\oplus\bm{4}\oplus\bm{5},~~\bm{4}\otimes\bm{4}=\bm{1_{S}}\oplus\bm{3_{A}}\oplus\bm{3^{\prime}_{A}}\oplus\bm{4_{S}}\oplus\bm{5_{S}},
        ~~\bm{4}\otimes\bm{5}=\bm{3}\oplus\bm{3^{\prime}}\oplus\bm{4}\oplus\bm{5_{1}}\oplus\bm{5_{2}},\\ \label{eq:Kronecker}&&\bm{5}\otimes\bm{5}=\bm{1_{S}}\oplus\bm{3_{A}}\oplus\bm{3^{\prime}_{A}}\oplus\bm{4_{S}}\oplus\bm{4_{A}}\oplus\bm{5_{S,1}}\oplus\bm{5_{S,2}}.
\end{eqnarray}
where $\bm{R}$ denotes any irreducible representation of $A_{5}$, and $\bm{4_{S}}$, $\bm{4_{A}}$,  $\bm{5_{1}}$, $\bm{5_{2}}$, $\bm{5_{S,1}}$ and  $\bm{5_{S,2}}$ stand for the two $\bm{4}$ and $\bm{5}$ which appear in the Kronecker products, where the subscript ``$\bm{S}$'' (``$\bm{A}$'') refers to symmetric (antisymmetric) combinations. We now list the CG coefficients in our basis, and the results are summarized in table~\ref{tab:A5_CG}.

\newpage

\begin{center}
\renewcommand{\arraystretch}{1.15}
\begin{small}
\setlength\LTcapwidth{\textwidth}
\setlength\LTleft{0.1in}
\setlength\LTright{0pt}
\begin{longtable}{|c|c|c|c|c|c|}
\caption{\label{tab:A5_CG}
Tensor products and the corresponding CG coefficients for the finite modular symmetry $A_{5}$. Here all CG coefficients are presented in the form $x\otimes y$, where $x_{i}$ denotes the elements of the left base vector $x$ and $y_{j}$ stands for the elements of the right base vector $y$.} \\
\midrule
\specialrule{0em}{1.0pt}{1.0pt}

\endfirsthead

\multicolumn{6}{c}
{{\bfseries \tablename\ \thetable{} -- continued from previous page}} \\
\hline

\endhead

\caption[]{continues on next page}\\
\endfoot

\endlastfoot

\hline
\multicolumn{2}{|c}{~$\bm{3}\otimes\bm{3}=\bm{1_{S}}\oplus\bm{3_{A}} \oplus\bm{5_{S}}$~} & \multicolumn{2}{|c}{ ~$\bm{3^\prime}\otimes\bm{3^\prime}=\bm{1_{S}}\oplus\bm{3^\prime_{A}} \oplus\bm{5_{S}}$~} & \multicolumn{2}{|c|}{~$\bm{3}\otimes\bm{3^\prime}=\bm{4}\oplus\bm{5}$~}  \\ \hline
                \multicolumn{2}{|c}{~} & \multicolumn{2}{|c}{~} & \multicolumn{2}{|c|}{~} \\[-0.18in]
\multicolumn{2}{|c}{~$\bm{1_S}:x_{1} y_{1}+x_{2} y_{3}+x_{3} y_{2}$~}  & \multicolumn{2}{|c}{~$\bm{1_{S}}: x_{1} y_{1}+x_{2} y_{3}+x_{3} y_{2}$~} & \multicolumn{2}{|c|}{~}  \\
\multicolumn{2}{|c}{~} & \multicolumn{2}{|c}{~} & \multicolumn{2}{|c|}{~} \\[-0.28in]
\multicolumn{2}{|c}{~$\bm{3_{A}}:\begin{pmatrix}
                        x_{2} y_{3}-x_{3} y_{2} \\
                        x_{1} y_{2}-x_{2} y_{1} \\
                        x_{3} y_{1}-x_{1} y_{3}
                \end{pmatrix}$~}
                 &
\multicolumn{2}{|c}{~$\bm{3^\prime_{A}}:
                \begin{pmatrix}
                        x_{2} y_{3}-x_{3} y_{2} \\
                        x_{1} y_{2}-x_{2} y_{1} \\
                        x_{3} y_{1}-x_{1} y_{3}
                \end{pmatrix}$~}
                ~ &
\multicolumn{2}{|c|}{~$\bm{4}:
                \begin{pmatrix}
                        \sqrt{2} x_{2} y_{1}+x_{3} y_{2} \\
                        -\sqrt{2} x_{1} y_{2}-x_{3} y_{3} \\
                        -\sqrt{2} x_{1} y_{3}-x_{2} y_{2} \\
                        \sqrt{2} x_{3} y_{1}+x_{2} y_{3}
                \end{pmatrix}$~} \\
\multicolumn{2}{|c}{~} & \multicolumn{2}{|c}{~} & \multicolumn{2}{|c|}{~} \\[-0.20in]
\multicolumn{2}{|c}{~$\bm{5_S}:
                \begin{pmatrix}
                        2 x_{1} y_{1}-x_{2} y_{3}-x_{3} y_{2} \\
                        -\sqrt{3} (x_{1} y_{2}+ x_{2} y_{1}) \\
                        \sqrt{6} x_{2} y_{2} \\
                        \sqrt{6} x_{3} y_{3} \\
                        -\sqrt{3} (x_{1} y_{3}+ x_{3} y_{1})
                \end{pmatrix}$~}
                ~  &
\multicolumn{2}{|c}{~$\bm{5_{S}}:
                \begin{pmatrix}
                        2 x_{1} y_{1}-x_{2} y_{3} -x_{3} y_{2}\\
                        \sqrt{6} x_{3} y_{3} \\
                        -\sqrt{3} (x_{1} y_{2} + x_{2} y_{1})\\
                        -\sqrt{3} (x_{1} y_{3}+ x_{3} y_{1}) \\
                        \sqrt{6} x_{2} y_{2}
                \end{pmatrix}$~}
                ~ &
\multicolumn{2}{|c|}{~$\bm{5}:
                \begin{pmatrix}
                        \sqrt{3} x_{1} y_{1} \\
                        x_{2} y_{1}-\sqrt{2} x_{3} y_{2} \\
                        x_{1} y_{2}-\sqrt{2} x_{3} y_{3} \\
                        x_{1} y_{3}-\sqrt{2} x_{2} y_{2} \\
                        x_{3} y_{1}-\sqrt{2} x_{2} y_{3}
                \end{pmatrix}$~}
                \\ \hline
\multicolumn{6}{c}{ } \\[-0.15in] \hline

\multicolumn{3}{|c}{~$\bm{3}\otimes\bm{4}=\bm{3^\prime}\oplus\bm{4}\oplus\bm{5}$~} & \multicolumn{3}{|c|}{~$\bm{3^\prime}\otimes\bm{4}=\bm{3}\oplus\bm{4}\oplus\bm{5}$~} \\ \hline

\multicolumn{3}{|c}{~}  & \multicolumn{3}{|c|}{~}\\[-0.16in]
\multicolumn{3}{|c}{~$\bm{3^\prime}:
                \begin{pmatrix}
                        -\sqrt{2} (x_{2} y_{4}+x_{3} y_{1}) \\
                        \sqrt{2} x_{1} y_{2}-x_{2} y_{1}+x_{3} y_{3} \\
                        \sqrt{2} x_{1} y_{3}+x_{2} y_{2}-x_{3} y_{4}
                \end{pmatrix}$~}
                 & \multicolumn{3}{|c|}{~$\bm{3}:
                \begin{pmatrix}
                        -\sqrt{2} (x_{2} y_{3}+ x_{3} y_{2}) \\
                        \sqrt{2} x_{1} y_{1}+x_{2} y_{4}-x_{3} y_{3} \\
                        \sqrt{2} x_{1} y_{4} -x_{2} y_{2}+x_{3} y_{1}
                \end{pmatrix}$~}\\
        \multicolumn{3}{|c}{~}  & \multicolumn{3}{|c|}{~} \\[-0.16in]
\multicolumn{3}{|c}{~$\bm{4}:
\begin{pmatrix}
        x_{1} y_{1}-\sqrt{2}x_{3} y_{2} \\
        -x_{1} y_{2}-\sqrt{2}x_{2} y_{1} \\
        x_{1} y_{3}+\sqrt{2}x_{3} y_{4} \\
        -x_{1} y_{4}+\sqrt{2}x_{2} y_{3}
\end{pmatrix}$~}
                 &
\multicolumn{3}{|c|}{~$\bm{4}:
                \begin{pmatrix}
                        x_{1} y_{1}+\sqrt{2}x_{3} y_{3} \\
                        x_{1} y_{2}-\sqrt{2}x_{3} y_{4} \\
                        -x_{1} y_{3}+\sqrt{2}x_{2} y_{1} \\
                        -x_{1} y_{4}-\sqrt{2}x_{2} y_{2}
                \end{pmatrix}$~}\\
\multicolumn{3}{|c}{~}  & \multicolumn{3}{|c|}{~}\\[-0.16in]
\multicolumn{3}{|c}{~$\bm{5}:
                \begin{pmatrix}
                        \sqrt{6} (x_{2}y_{4}- x_{3} y_{1}) \\
                        2\sqrt{2} x_{1} y_{1}+2 x_{3} y_{2}\\
                        -\sqrt{2} x_{1} y_{2}+x_{2} y_{1}+3x_{3} y_{3} \\
                        \sqrt{2} x_{1} y_{3}-3x_{2} y_{2}-x_{3} y_{4}\\
                        -2\sqrt{2} x_{1} y_{4}-2 x_{2} y_{3}
                \end{pmatrix}$~}
                 &
\multicolumn{3}{|c|}{~$\bm{5}:
                \begin{pmatrix}
                        \sqrt{6} (x_{2} y_{3}-x_{3}y_{2}) \\
                        \sqrt{2} x_{1} y_{1}-3x_{2} y_{4}-x_{3} y_{3} \\
                        2\sqrt{2} x_{1} y_{2}+2 x_{3} y_{4} \\
                        -2\sqrt{2} x_{1} y_{3}-2x_{2} y_{1} \\
                        -\sqrt{2}x_{1} y_{4}+x_{2} y_{2}+3x_{3} y_{1}
                \end{pmatrix}$~} \\ \hline 

\multicolumn{6}{c}{ } \\[-0.15in] \hline

\multicolumn{3}{|c}{~$\bm{3}\otimes\bm{5}=\bm{3}\oplus\bm{3^\prime}\oplus\bm{4}\oplus\bm{5}$~} & \multicolumn{3}{|c|}{~$\bm{3^\prime}\otimes\bm{5}=\bm{3}\oplus\bm{3^\prime}\oplus\bm{4}\oplus\bm{5}$~} \\ \hline
\multicolumn{3}{|c}{~}  & \multicolumn{3}{|c|}{~}\\[-0.16in]
\multicolumn{3}{|c}{~$\bm{3}:
\begin{pmatrix}
        -2 x_{1} y_{1}+\sqrt{3}x_{2} y_{5}+\sqrt{3}x_{3} y_{2} \\
        \sqrt{3}x_{1} y_{2}+x_{2} y_{1}-\sqrt{6}x_{3} y_{3} \\
        \sqrt{3}x_{1} y_{5}-\sqrt{6}x_{2} y_{4}+x_{3} y_{1}
\end{pmatrix}$~}
 &
\multicolumn{3}{|c|}{~$\bm{3}:
\begin{pmatrix}
        \sqrt{3} x_{1} y_{1}+x_{2}y_{4}+x_{3} y_{3} \\
        x_{1} y_{2}-\sqrt{2}x_{2} y_{5} -\sqrt{2}x_{3} y_{4}\\
        x_{1} y_{5}-\sqrt{2} x_{2} y_{3}-\sqrt{2} x_{3} y_{2}
\end{pmatrix}$~} \\
\multicolumn{3}{|c}{~}  & \multicolumn{3}{|c|}{~}\\[-0.16in]
\multicolumn{3}{|c}{~$\bm{3^\prime}:
\begin{pmatrix}
        \sqrt{3} x_{1} y_{1}+x_{2} y_{5}+x_{3} y_{2} \\
        x_{1} y_{3}-\sqrt{2}x_{2} y_{2}-\sqrt{2}x_{3} y_{4} \\
        x_{1} y_{4}-\sqrt{2}x_{2} y_{3}-\sqrt{2}x_{3} y_{5}
\end{pmatrix}$~}
 & \multicolumn{3}{|c|}{~$\bm{3^\prime}:
\begin{pmatrix}
        -2 x_{1} y_{1}+\sqrt{3}x_{2} y_{4} +\sqrt{3}x_{3} y_{3}\\
        \sqrt{3}x_{1} y_{3}+x_{2} y_{1}-\sqrt{6}x_{3} y_{5} \\
        \sqrt{3}x_{1} y_{4}-\sqrt{6}x_{2} y_{2}+x_{3} y_{1}
\end{pmatrix}$~} \\
  \multicolumn{3}{|c}{~}        & \multicolumn{3}{|c|}{~}\\[-0.16in]
\multicolumn{3}{|c}{~$\bm{4}:
\begin{pmatrix}
        2\sqrt{2} x_{1} y_{2}-\sqrt{6} x_{2} y_{1}+x_{3} y_{3} \\
        -\sqrt{2}x_{1} y_{3}+2x_{2} y_{2}-3 x_{3} y_{4} \\
        \sqrt{2}x_{1} y_{4}+3x_{2} y_{3}-2x_{3} y_{5} \\
        -2\sqrt{2} x_{1} y_{5}-x_{2} y_{4}+\sqrt{6} x_{3} y_{1}
\end{pmatrix}$~}
 &  \multicolumn{3}{|c|}{~$\bm{4}:
\begin{pmatrix}
        \sqrt{2} x_{1} y_{2}+3 x_{2}y_{5}-2x_{3} y_{4} \\
        2\sqrt{2} x_{1} y_{3}-\sqrt{6} x_{2} y_{1}+x_{3} y_{5} \\
        -2\sqrt{2} x_{1} y_{4}-x_{2} y_{2} +\sqrt{6} x_{3} y_{1}\\
        -\sqrt{2} x_{1}y_{5}+2 x_{2}y_{3}-3 x_{3} y_{2}
\end{pmatrix}$~}
 \\
\multicolumn{3}{|c}{~}  & \multicolumn{3}{|c|}{~}\\[-0.16in]
\multicolumn{3}{|c}{~$\bm{5}:
\begin{pmatrix}
        \sqrt{3} (x_{2} y_{5}- x_{3}y_{2}) \\
        -x_{1} y_{2}-\sqrt{3} x_{2} y_{1}-\sqrt{2}x_{3} y_{3} \\
        -2 x_{1} y_{3}-\sqrt{2}x_{2} y_{2} \\
        2x_{1} y_{4}+\sqrt{2}x_{3} y_{5} \\
        x_{1} y_{5}+\sqrt{2}x_{2} y_{4}+ \sqrt{3} x_{3} y_{1}
\end{pmatrix}$~}
 &
\multicolumn{3}{|c|}{~$\bm{5}:
\begin{pmatrix}
        \sqrt{3} (x_{2} y_{4}- x_{3} y_{3}) \\
        2 x_{1} y_{2}+\sqrt{2}x_{3} y_{4} \\
        -x_{1} y_{3}-\sqrt{3} x_{2} y_{1}-\sqrt{2}x_{3} y_{5} \\
        x_{1} y_{4}+\sqrt{2} x_{2} y_{2} + \sqrt{3} x_{3} y_{1}\\
        -2x_{1} y_{5}-\sqrt{2} x_{2} y_{3}
\end{pmatrix}$~}
 \\ \hline

\multicolumn{6}{c}{ } \\[-0.15in] \hline

\multicolumn{6}{|c|}{$\bm{5}\otimes\bm{5}=\bm{1_{S}}\oplus\bm{3_{A}}\oplus\bm{3^\prime_{A}}\oplus\bm{4_{S}}\oplus\bm{4_{A}}\oplus\bm{5_{S,1}}\oplus\bm{5_{S,2}}$}  \\ \hline
\multicolumn{6}{|c|}{  $ \begin{array}{l}
\bm{1_{S}}:x_{1}y_{1}+x_{2}y_{5}+x_{3}y_{4}+x_{4}y_{3}+x_{5}y_{2}~ \\
                ~\bm{3_{A}}:
                \begin{pmatrix}
                        x_{2} y_{5}+2x_{3} y_{4}-2 x_{4} y_{3}-x_{5} y_{2} \\
                        -\sqrt{3} x_{1}y_{2}+\sqrt{3} x_{2} y_{1}+\sqrt{2}x_{3} y_{5}-\sqrt{2} x_{5} y_{3} \\
                        \sqrt{3}x_{1} y_{5}+\sqrt{2} x_{2} y_{4}-\sqrt{2} x_{4}y_{2}-\sqrt{3} x_{5} y_{1}
                \end{pmatrix}
                ~ \\
                ~\bm{3^\prime_{A}}:
                \begin{pmatrix}
                        2 x_{2} y_{5}-x_{3} y_{4}+x_{4} y_{3}-2 x_{5} y_{2}\\
                        \sqrt{3} x_{1}y_{3}-\sqrt{3} x_{3} y_{1}+\sqrt{2}x_{4} y_{5}-\sqrt{2} x_{5} y_{4} \\
                        -\sqrt{3}x_{1} y_{4}+\sqrt{2} x_{2} y_{3}-\sqrt{2} x_{3}y_{2}+\sqrt{3} x_{4} y_{1}
                \end{pmatrix}
                ~ \\
                ~\bm{4_{S}}:
                \begin{pmatrix}
                        3 \sqrt{2} x_{1} y_{2}+3 \sqrt{2} x_{2} y_{1}-\sqrt{3}x_{3}y_{5}+4 \sqrt{3} x_{4} y_{4}-\sqrt{3}x_{5} y_{3} \\
                        3 \sqrt{2}x_{1} y_{3}+4 \sqrt{3} x_{2} y_{2}+3 \sqrt{2} x_{3} y_{1}-\sqrt{3}x_{4}y_{5}-\sqrt{3} x_{5} y_{4} \\
                        3 \sqrt{2} x_{1} y_{4}-\sqrt{3}x_{2} y_{3}-\sqrt{3} x_{3} y_{2}+3 \sqrt{2} x_{4} y_{1}+4 \sqrt{3}x_{5}y_{5} \\
                        3 \sqrt{2} x_{1}y_{5}-\sqrt{3} x_{2} y_{4}+4 \sqrt{3}x_{3} y_{3}-\sqrt{3} x_{4} y_{2}+3 \sqrt{2} x_{5} y_{1}
                \end{pmatrix}
                ~ \\
                ~\bm{4_{A}}:
                \begin{pmatrix}
                        \sqrt{2} x_{1} y_{2}-\sqrt{2} x_{2} y_{1}+\sqrt{3} x_{3} y_{5}-\sqrt{3}x_{5} y_{3} \\
                        -\sqrt{2} x_{1} y_{3}+\sqrt{2} x_{3} y_{1}+\sqrt{3} x_{4} y_{5}-\sqrt{3} x_{5} y_{4} \\
                        -\sqrt{2} x_{1} y_{4}-\sqrt{3} x_{2} y_{3}+\sqrt{3} x_{3} y_{2}+\sqrt{2} x_{4} y_{1}\\
                        \sqrt{2} x_{1} y_{5}-\sqrt{3}x_{2} y_{4}+\sqrt{3} x_{4} y_{2}-\sqrt{2} x_{5} y_{1}
                \end{pmatrix}
                ~  \\
                ~\bm{5_{S,1}}:
                \begin{pmatrix}
                        2 x_{1} y_{1}+x_{2} y_{5}-2 x_{3} y_{4}-2 x_{4} y_{3}+x_{5} y_{2} \\
                        x_{1} y_{2}+x_{2} y_{1}+\sqrt{6} x_{3} y_{5}+\sqrt{6} x_{5} y_{3} \\
                        -2 x_{1} y_{3}+\sqrt{6} x_{2} y_{2}-2 x_{3} y_{1} \\
                        -2 x_{1} y_{4}-2 x_{4} y_{1}+\sqrt{6} x_{5} y_{5} \\
                        x_{1} y_{5}+\sqrt{6} x_{2} y_{4}+\sqrt{6} x_{4} y_{2}+x_{5} y_{1}
                \end{pmatrix}
                ~ \\
                ~\bm{5_{S,2}}:
                \begin{pmatrix}
                        2 x_{1} y_{1}-2 x_{2} y_{5}+x_{3} y_{4}+x_{4}y_{3}-2 x_{5} y_{2} \\
                        -2 x_{1} y_{2}-2 x_{2} y_{1}+\sqrt{6} x_{4} y_{4} \\
                        x_{1} y_{3}+x_{3} y_{1}+\sqrt{6} x_{4} y_{5}+\sqrt{6} x_{5} y_{4} \\
                        x_{1} y_{4}+\sqrt{6} x_{2} y_{3}+\sqrt{6} x_{3} y_{2}+x_{4} y_{1} \\
                        -2 x_{1} y_{5}+\sqrt{6} x_{3} y_{3}-2 x_{5} y_{1}
                \end{pmatrix}
\\
\end{array} $ } \\
\hline

\specialrule{0em}{1.0pt}{1.0pt}
\midrule
\end{longtable}
\end{small}
\end{center}

\end{appendix}



\begin{thebibliography}{10}

\bibitem{ParticleDataGroup:2024cfk}
{\bfseries Particle Data Group} Collaboration, S.~Navas {\em et~al.}, ``{Review
  of particle physics},''
  \href{http://dx.doi.org/10.1103/PhysRevD.110.030001}{{\em Phys. Rev. D}
  {\bfseries 110} no.~3, (2024) 030001}.

\bibitem{King:2017guk}
S.~F. King, ``{Unified Models of Neutrinos, Flavour and CP Violation},''
  \href{http://dx.doi.org/10.1016/j.ppnp.2017.01.003}{{\em Prog. Part. Nucl.
  Phys.} {\bfseries 94} (2017) 217--256},
  \href{http://arxiv.org/abs/1701.04413}{{\ttfamily arXiv:1701.04413
  [hep-ph]}}.

\bibitem{Petcov:2017ggy}
S.~T. Petcov, ``{Discrete Flavour Symmetries, Neutrino Mixing and Leptonic CP
  Violation},'' \href{http://dx.doi.org/10.1140/epjc/s10052-018-6158-5}{{\em
  Eur. Phys. J. C} {\bfseries 78} no.~9, (2018) 709},
  \href{http://arxiv.org/abs/1711.10806}{{\ttfamily arXiv:1711.10806
  [hep-ph]}}.

\bibitem{Xing:2020ijf}
Z.-z. Xing, ``{Flavor structures of charged fermions and massive neutrinos},''
  \href{http://dx.doi.org/10.1016/j.physrep.2020.02.001}{{\em Phys. Rept.}
  {\bfseries 854} (2020) 1--147},
  \href{http://arxiv.org/abs/1909.09610}{{\ttfamily arXiv:1909.09610
  [hep-ph]}}.

\bibitem{Feruglio:2019ybq}
F.~Feruglio and A.~Romanino, ``{Lepton flavor symmetries},''
  \href{http://dx.doi.org/10.1103/RevModPhys.93.015007}{{\em Rev. Mod. Phys.}
  {\bfseries 93} no.~1, (2021) 015007},
  \href{http://arxiv.org/abs/1912.06028}{{\ttfamily arXiv:1912.06028
  [hep-ph]}}.

\bibitem{Ding:2024ozt}
G.-J. Ding and J.~W.~F. Valle, ``{The symmetry approach to quark and lepton
  masses and mixing},'' \href{http://arxiv.org/abs/2402.16963}{{\ttfamily
  arXiv:2402.16963 [hep-ph]}}.

\bibitem{Feruglio:2017spp}
F.~Feruglio, {\em {Are neutrino masses modular forms?}},
  \href{http://dx.doi.org/10.1142/9789813238053_0012}{pp.~227--266}.
\newblock 2019.
\newblock \href{http://arxiv.org/abs/1706.08749}{{\ttfamily arXiv:1706.08749
  [hep-ph]}}.

\bibitem{Lauer:1989ax}
J.~Lauer, J.~Mas, and H.~P. Nilles, ``{Duality and the Role of Nonperturbative
  Effects on the World Sheet},''
  \href{http://dx.doi.org/10.1016/0370-2693(89)91190-8}{{\em Phys. Lett. B}
  {\bfseries 226} (1989) 251--256}.

\bibitem{Ferrara:1989bc}
S.~Ferrara, D.~Lust, A.~D. Shapere, and S.~Theisen, ``{Modular Invariance in
  Supersymmetric Field Theories},''
  \href{http://dx.doi.org/10.1016/0370-2693(89)90583-2}{{\em Phys. Lett. B}
  {\bfseries 225} (1989) 363}.

\bibitem{Ferrara:1989qb}
S.~Ferrara, .~D. Lust, and S.~Theisen, ``{Target Space Modular Invariance and
  Low-Energy Couplings in Orbifold Compactifications},''
  \href{http://dx.doi.org/10.1016/0370-2693(89)90631-X}{{\em Phys. Lett. B}
  {\bfseries 233} (1989) 147--152}.

\bibitem{deAdelhartToorop:2011re}
R.~de~Adelhart~Toorop, F.~Feruglio, and C.~Hagedorn, ``{Finite Modular Groups
  and Lepton Mixing},''
  \href{http://dx.doi.org/10.1016/j.nuclphysb.2012.01.017}{{\em Nucl. Phys. B}
  {\bfseries 858} (2012) 437--467},
  \href{http://arxiv.org/abs/1112.1340}{{\ttfamily arXiv:1112.1340 [hep-ph]}}.

\bibitem{Kobayashi:2023zzc}
T.~Kobayashi and M.~Tanimoto, ``{Modular flavor symmetric models},''
\newblock 7, 2023.
\newblock \href{http://arxiv.org/abs/2307.03384}{{\ttfamily arXiv:2307.03384
  [hep-ph]}}.

\bibitem{Ding:2023htn}
G.-J. Ding and S.~F. King, ``{Neutrino mass and mixing with modular
  symmetry},'' \href{http://dx.doi.org/10.1088/1361-6633/ad52a3}{{\em Rept.
  Prog. Phys.} {\bfseries 87} no.~8, (2024) 084201},
  \href{http://arxiv.org/abs/2311.09282}{{\ttfamily arXiv:2311.09282
  [hep-ph]}}.

\bibitem{Ding:2022nzn}
G.-J. Ding, X.-G. Liu, and C.-Y. Yao, ``{A minimal modular invariant neutrino
  model},'' \href{http://dx.doi.org/10.1007/JHEP01(2023)125}{{\em JHEP}
  {\bfseries 01} (2023) 125}, \href{http://arxiv.org/abs/2211.04546}{{\ttfamily
  arXiv:2211.04546 [hep-ph]}}.

\bibitem{Ding:2023ydy}
G.-J. Ding, X.-G. Liu, J.-N. Lu, and M.-H. Weng, ``{Modular binary octahedral
  symmetry for flavor structure of Standard Model},''
  \href{http://dx.doi.org/10.1007/JHEP11(2023)083}{{\em JHEP} {\bfseries 11}
  (2023) 083}, \href{http://arxiv.org/abs/2307.14926}{{\ttfamily
  arXiv:2307.14926 [hep-ph]}}.

\bibitem{Ding:2024pix}
G.-J. Ding, E.~Lisi, A.~Marrone, and S.~T. Petcov, ``{Interplay and
  Correlations Between Quark and Lepton Observables in Modular Symmetry
  Models},'' \href{http://arxiv.org/abs/2409.15823}{{\ttfamily arXiv:2409.15823
  [hep-ph]}}.

\bibitem{Feruglio:2022koo}
F.~Feruglio, ``{Universal Predictions of Modular Invariant Flavor Models near
  the Self-Dual Point},''
  \href{http://dx.doi.org/10.1103/PhysRevLett.130.101801}{{\em Phys. Rev.
  Lett.} {\bfseries 130} no.~10, (2023) 101801},
  \href{http://arxiv.org/abs/2211.00659}{{\ttfamily arXiv:2211.00659
  [hep-ph]}}.

\bibitem{Feruglio:2023mii}
F.~Feruglio, ``{Fermion masses, critical behavior and universality},''
  \href{http://dx.doi.org/10.1007/JHEP03(2023)236}{{\em JHEP} {\bfseries 03}
  (2023) 236}, \href{http://arxiv.org/abs/2302.11580}{{\ttfamily
  arXiv:2302.11580 [hep-ph]}}.

\bibitem{Ding:2024xhz}
G.-J. Ding, F.~Feruglio, and X.-G. Liu, ``{Universal predictions of Siegel
  modular invariant theories near the fixed points},''
  \href{http://dx.doi.org/10.1007/JHEP05(2024)052}{{\em JHEP} {\bfseries 05}
  (2024) 052}, \href{http://arxiv.org/abs/2402.14915}{{\ttfamily
  arXiv:2402.14915 [hep-ph]}}.

\bibitem{Okada:2020ukr}
H.~Okada and M.~Tanimoto, ``{Modular invariant flavor model of $A_4$ and
  hierarchical structures at nearby fixed points},''
  \href{http://dx.doi.org/10.1103/PhysRevD.103.015005}{{\em Phys. Rev. D}
  {\bfseries 103} no.~1, (2021) 015005},
  \href{http://arxiv.org/abs/2009.14242}{{\ttfamily arXiv:2009.14242
  [hep-ph]}}.

\bibitem{Feruglio:2021dte}
F.~Feruglio, V.~Gherardi, A.~Romanino, and A.~Titov, ``{Modular invariant
  dynamics and fermion mass hierarchies around $\tau = i$},''
  \href{http://dx.doi.org/10.1007/JHEP05(2021)242}{{\em JHEP} {\bfseries 05}
  (2021) 242}, \href{http://arxiv.org/abs/2101.08718}{{\ttfamily
  arXiv:2101.08718 [hep-ph]}}.

\bibitem{Novichkov:2021evw}
P.~P. Novichkov, J.~T. Penedo, and S.~T. Petcov, ``{Fermion mass hierarchies,
  large lepton mixing and residual modular symmetries},''
  \href{http://dx.doi.org/10.1007/JHEP04(2021)206}{{\em JHEP} {\bfseries 04}
  (2021) 206}, \href{http://arxiv.org/abs/2102.07488}{{\ttfamily
  arXiv:2102.07488 [hep-ph]}}.

\bibitem{Petcov:2022fjf}
S.~T. Petcov and M.~Tanimoto, ``{$A_4$ modular flavour model of quark mass
  hierarchies close to the fixed point $\tau = \omega $},''
  \href{http://dx.doi.org/10.1140/epjc/s10052-023-11727-0}{{\em Eur. Phys. J.
  C} {\bfseries 83} no.~7, (2023) 579},
  \href{http://arxiv.org/abs/2212.13336}{{\ttfamily arXiv:2212.13336
  [hep-ph]}}.

\bibitem{Kikuchi:2023cap}
S.~Kikuchi, T.~Kobayashi, K.~Nasu, S.~Takada, and H.~Uchida, ``{Quark
  hierarchical structures in modular symmetric flavor models at level 6},''
  \href{http://dx.doi.org/10.1103/PhysRevD.107.055014}{{\em Phys. Rev. D}
  {\bfseries 107} no.~5, (2023) 055014},
  \href{http://arxiv.org/abs/2301.03737}{{\ttfamily arXiv:2301.03737
  [hep-ph]}}.

\bibitem{Abe:2023ilq}
Y.~Abe, T.~Higaki, J.~Kawamura, and T.~Kobayashi, ``{Quark masses and CKM
  hierarchies from $S_4'$ modular flavor symmetry},''
  \href{http://dx.doi.org/10.1140/epjc/s10052-023-12303-2}{{\em Eur. Phys. J.
  C} {\bfseries 83} no.~12, (2023) 1140},
  \href{http://arxiv.org/abs/2301.07439}{{\ttfamily arXiv:2301.07439
  [hep-ph]}}.

\bibitem{Kikuchi:2023jap}
S.~Kikuchi, T.~Kobayashi, K.~Nasu, S.~Takada, and H.~Uchida, ``{Quark mass
  hierarchies and CP violation in A$_{4}$ \texttimes{} A$_{4}$ \texttimes{}
  A$_{4}$ modular symmetric flavor models},''
  \href{http://dx.doi.org/10.1007/JHEP07(2023)134}{{\em JHEP} {\bfseries 07}
  (2023) 134}, \href{http://arxiv.org/abs/2302.03326}{{\ttfamily
  arXiv:2302.03326 [hep-ph]}}.

\bibitem{Abe:2023qmr}
Y.~Abe, T.~Higaki, J.~Kawamura, and T.~Kobayashi, ``{Quark and lepton
  hierarchies from S4' modular flavor symmetry},''
  \href{http://dx.doi.org/10.1016/j.physletb.2023.137977}{{\em Phys. Lett. B}
  {\bfseries 842} (2023) 137977},
  \href{http://arxiv.org/abs/2302.11183}{{\ttfamily arXiv:2302.11183
  [hep-ph]}}.

\bibitem{Petcov:2023vws}
S.~T. Petcov and M.~Tanimoto, ``{A$_{4}$ modular flavour model of quark mass
  hierarchies close to the fixed point \ensuremath{\tau} =
  i\ensuremath{\infty}},''
  \href{http://dx.doi.org/10.1007/JHEP08(2023)086}{{\em JHEP} {\bfseries 08}
  (2023) 086}, \href{http://arxiv.org/abs/2306.05730}{{\ttfamily
  arXiv:2306.05730 [hep-ph]}}.

\bibitem{deMedeirosVarzielas:2023crv}
I.~de~Medeiros~Varzielas, M.~Levy, J.~T. Penedo, and S.~T. Petcov, ``{Quarks at
  the modular S$_{4}$ cusp},''
  \href{http://dx.doi.org/10.1007/JHEP09(2023)196}{{\em JHEP} {\bfseries 09}
  (2023) 196}, \href{http://arxiv.org/abs/2307.14410}{{\ttfamily
  arXiv:2307.14410 [hep-ph]}}.

\bibitem{Feruglio:2023uof}
F.~Feruglio, A.~Strumia, and A.~Titov, ``{Modular invariance and the QCD
  angle},'' \href{http://dx.doi.org/10.1007/JHEP07(2023)027}{{\em JHEP}
  {\bfseries 07} (2023) 027}, \href{http://arxiv.org/abs/2305.08908}{{\ttfamily
  arXiv:2305.08908 [hep-ph]}}.

\bibitem{Petcov:2024vph}
S.~T. Petcov and M.~Tanimoto, ``{$A_4$ modular invariance and the strong CP
  problem},'' \href{http://dx.doi.org/10.1140/epjc/s10052-024-13272-w}{{\em
  Eur. Phys. J. C} {\bfseries 84} no.~9, (2024) 914},
  \href{http://arxiv.org/abs/2404.00858}{{\ttfamily arXiv:2404.00858
  [hep-ph]}}.

\bibitem{Penedo:2024gtb}
J.~T. Penedo and S.~T. Petcov, ``{Finite modular symmetries and the strong CP
  problem},'' \href{http://arxiv.org/abs/2404.08032}{{\ttfamily
  arXiv:2404.08032 [hep-ph]}}.

\bibitem{Feruglio:2024ytl}
F.~Feruglio, M.~Parriciatu, A.~Strumia, and A.~Titov, ``{Solving the strong CP
  problem without axions},''
  \href{http://dx.doi.org/10.1007/JHEP08(2024)214}{{\em JHEP} {\bfseries 08}
  (2024) 214}, \href{http://arxiv.org/abs/2406.01689}{{\ttfamily
  arXiv:2406.01689 [hep-ph]}}.

\bibitem{Ding:2024neh}
G.-J. Ding, S.-Y. Jiang, and W.~Zhao, ``{Modular Invariant Slow Roll
  Inflation},'' \href{http://arxiv.org/abs/2405.06497}{{\ttfamily
  arXiv:2405.06497 [hep-ph]}}.

\bibitem{King:2024ssx}
S.~F. King and X.~Wang, ``{Modular invariant hilltop inflation},''
  \href{http://dx.doi.org/10.1088/1475-7516/2024/07/073}{{\em JCAP} {\bfseries
  07} (2024) 073}, \href{http://arxiv.org/abs/2405.08924}{{\ttfamily
  arXiv:2405.08924 [hep-ph]}}.

\bibitem{Baur:2019kwi}
A.~Baur, H.~P. Nilles, A.~Trautner, and P.~K.~S. Vaudrevange, ``{Unification of
  Flavor, CP, and Modular Symmetries},''
  \href{http://dx.doi.org/10.1016/j.physletb.2019.03.066}{{\em Phys. Lett. B}
  {\bfseries 795} (2019) 7--14},
  \href{http://arxiv.org/abs/1901.03251}{{\ttfamily arXiv:1901.03251
  [hep-th]}}.

\bibitem{Nilles:2020nnc}
H.~P. Nilles, S.~Ramos-S\'anchez, and P.~K.~S. Vaudrevange, ``{Eclectic Flavor
  Groups},'' \href{http://dx.doi.org/10.1007/JHEP02(2020)045}{{\em JHEP}
  {\bfseries 02} (2020) 045}, \href{http://arxiv.org/abs/2001.01736}{{\ttfamily
  arXiv:2001.01736 [hep-ph]}}.

\bibitem{Nilles:2020kgo}
H.~P. Nilles, S.~Ramos-Sanchez, and P.~K.~S. Vaudrevange, ``{Lessons from
  eclectic flavor symmetries},''
  \href{http://dx.doi.org/10.1016/j.nuclphysb.2020.115098}{{\em Nucl. Phys. B}
  {\bfseries 957} (2020) 115098},
  \href{http://arxiv.org/abs/2004.05200}{{\ttfamily arXiv:2004.05200
  [hep-ph]}}.

\bibitem{Nilles:2020tdp}
H.~P. Nilles, S.~Ramos\textendash{}S\'anchez, and P.~K.~S. Vaudrevange,
  ``{Eclectic flavor scheme from ten-dimensional string theory \textendash{} I.
  Basic results},''
  \href{http://dx.doi.org/10.1016/j.physletb.2020.135615}{{\em Phys. Lett. B}
  {\bfseries 808} (2020) 135615},
  \href{http://arxiv.org/abs/2006.03059}{{\ttfamily arXiv:2006.03059
  [hep-th]}}.

\bibitem{Baur:2020jwc}
A.~Baur, M.~Kade, H.~P. Nilles, S.~Ramos-Sanchez, and P.~K.~S. Vaudrevange,
  ``{The eclectic flavor symmetry of the $\boldsymbol{\mathbb{Z}_2}$
  orbifold},'' \href{http://dx.doi.org/10.1007/JHEP02(2021)018}{{\em JHEP}
  {\bfseries 02} (2021) 018}, \href{http://arxiv.org/abs/2008.07534}{{\ttfamily
  arXiv:2008.07534 [hep-th]}}.

\bibitem{Nilles:2020gvu}
H.~P. Nilles, S.~Ramos\textendash{}S\'anchez, and P.~K.~S. Vaudrevange,
  ``{Eclectic flavor scheme from ten-dimensional string theory - II detailed
  technical analysis},''
  \href{http://dx.doi.org/10.1016/j.nuclphysb.2021.115367}{{\em Nucl. Phys. B}
  {\bfseries 966} (2021) 115367},
  \href{http://arxiv.org/abs/2010.13798}{{\ttfamily arXiv:2010.13798
  [hep-th]}}.

\bibitem{Baur:2021mtl}
A.~Baur, M.~Kade, H.~P. Nilles, S.~Ramos-Sanchez, and P.~K.~S. Vaudrevange,
  ``{Completing the eclectic flavor scheme of the \ensuremath{\mathbb{Z}}$_{2}$
  orbifold},'' \href{http://dx.doi.org/10.1007/JHEP06(2021)110}{{\em JHEP}
  {\bfseries 06} (2021) 110}, \href{http://arxiv.org/abs/2104.03981}{{\ttfamily
  arXiv:2104.03981 [hep-th]}}.

\bibitem{Nilles:2023shk}
H.~P. Nilles and S.~Ramos-Sanchez, ``{The flavor puzzle: Textures and
  symmetries},'' \href{http://dx.doi.org/10.1142/S0217751X24410033}{{\em Int.
  J. Mod. Phys. A} {\bfseries 39} no.~09n10, (2024) 2441003},
  \href{http://arxiv.org/abs/2308.14810}{{\ttfamily arXiv:2308.14810
  [hep-ph]}}.

\bibitem{Chen:2019ewa}
M.-C. Chen, S.~Ramos-S\'anchez, and M.~Ratz, ``{A note on the predictions of
  models with modular flavor symmetries},''
  \href{http://dx.doi.org/10.1016/j.physletb.2019.135153}{{\em Phys. Lett. B}
  {\bfseries 801} (2020) 135153},
  \href{http://arxiv.org/abs/1909.06910}{{\ttfamily arXiv:1909.06910
  [hep-ph]}}.

\bibitem{Chen:2021prl}
M.-C. Chen, V.~Knapp-Perez, M.~Ramos-Hamud, S.~Ramos-Sanchez, M.~Ratz, and
  S.~Shukla, ``{Quasi\textendash{}eclectic modular flavor symmetries},''
  \href{http://dx.doi.org/10.1016/j.physletb.2021.136843}{{\em Phys. Lett. B}
  {\bfseries 824} (2022) 136843},
  \href{http://arxiv.org/abs/2108.02240}{{\ttfamily arXiv:2108.02240
  [hep-ph]}}.

\bibitem{Baur:2022hma}
A.~Baur, H.~P. Nilles, S.~Ramos-Sanchez, A.~Trautner, and P.~K.~S. Vaudrevange,
  ``{The first string-derived eclectic flavor model with realistic
  phenomenology},'' \href{http://dx.doi.org/10.1007/JHEP09(2022)224}{{\em JHEP}
  {\bfseries 09} (2022) 224}, \href{http://arxiv.org/abs/2207.10677}{{\ttfamily
  arXiv:2207.10677 [hep-ph]}}.

\bibitem{Ding:2023ynd}
G.-J. Ding, S.~F. King, C.-C. Li, X.-G. Liu, and J.-N. Lu, ``{Neutrino mass and
  mixing models with eclectic flavor symmetry \ensuremath{\Delta}(27)
  \ensuremath{\rtimes} T'},''
  \href{http://dx.doi.org/10.1007/JHEP05(2023)144}{{\em JHEP} {\bfseries 05}
  (2023) 144}, \href{http://arxiv.org/abs/2303.02071}{{\ttfamily
  arXiv:2303.02071 [hep-ph]}}.

\bibitem{Li:2023dvm}
C.-C. Li and G.-J. Ding, ``{Eclectic flavor group \ensuremath{\Delta}(27)
  \ensuremath{\rtimes} S$_{3}$ and lepton model building},''
  \href{http://dx.doi.org/10.1007/JHEP03(2024)054}{{\em JHEP} {\bfseries 03}
  (2024) 054}, \href{http://arxiv.org/abs/2308.16901}{{\ttfamily
  arXiv:2308.16901 [hep-ph]}}.

\bibitem{Li:2024pff}
C.-C. Li, J.-N. Lu, and G.-J. Ding, ``{Minimal eclectic flavor group
  $Q_{8}\rtimes S_3$ and neutrino mixing},''
  \href{http://arxiv.org/abs/2405.13460}{{\ttfamily arXiv:2405.13460
  [hep-ph]}}.

\bibitem{Qu:2024rns}
B.-Y. Qu and G.-J. Ding, ``{Non-holomorphic modular flavor symmetry},''
  \href{http://dx.doi.org/10.1007/JHEP08(2024)136}{{\em JHEP} {\bfseries 08}
  (2024) 136}, \href{http://arxiv.org/abs/2406.02527}{{\ttfamily
  arXiv:2406.02527 [hep-ph]}}.

\bibitem{Novichkov:2019sqv}
P.~P. Novichkov, J.~T. Penedo, S.~T. Petcov, and A.~V. Titov, ``{Generalised CP
  Symmetry in Modular-Invariant Models of Flavour},''
  \href{http://dx.doi.org/10.1007/JHEP07(2019)165}{{\em JHEP} {\bfseries 07}
  (2019) 165}, \href{http://arxiv.org/abs/1905.11970}{{\ttfamily
  arXiv:1905.11970 [hep-ph]}}.

\bibitem{Nomura:2024atp}
T.~Nomura and H.~Okada, ``{Type-II seesaw of a non-holomorphic modular $A_4$
  symmetry},'' \href{http://arxiv.org/abs/2408.01143}{{\ttfamily
  arXiv:2408.01143 [hep-ph]}}.

\bibitem{Ding:2024inn}
G.-J. Ding, J.-N. Lu, S.~T. Petcov, and B.-Y. Qu, ``{Non-holomorphic Modular
  $S_4$ Lepton Flavour Models},''
  \href{http://arxiv.org/abs/2408.15988}{{\ttfamily arXiv:2408.15988
  [hep-ph]}}.

\bibitem{Liu:2019khw}
X.-G. Liu and G.-J. Ding, ``{Neutrino Masses and Mixing from Double Covering of
  Finite Modular Groups},''
  \href{http://dx.doi.org/10.1007/JHEP08(2019)134}{{\em JHEP} {\bfseries 08}
  (2019) 134}, \href{http://arxiv.org/abs/1907.01488}{{\ttfamily
  arXiv:1907.01488 [hep-ph]}}.

\bibitem{Esteban:2020cvm}
I.~Esteban, M.~C. Gonzalez-Garcia, M.~Maltoni, T.~Schwetz, and A.~Zhou, ``{The
  fate of hints: updated global analysis of three-flavor neutrino
  oscillations},'' \href{http://dx.doi.org/10.1007/JHEP09(2020)178}{{\em JHEP}
  {\bfseries 09} (2020) 178}, \href{http://arxiv.org/abs/2007.14792}{{\ttfamily
  arXiv:2007.14792 [hep-ph]}}.

\bibitem{JUNO:2022mxj}
{\bfseries JUNO} Collaboration, A.~Abusleme {\em et~al.}, ``{Sub-percent
  precision measurement of neutrino oscillation parameters with JUNO},''
  \href{http://dx.doi.org/10.1088/1674-1137/ac8bc9}{{\em Chin. Phys. C}
  {\bfseries 46} no.~12, (2022) 123001},
  \href{http://arxiv.org/abs/2204.13249}{{\ttfamily arXiv:2204.13249
  [hep-ex]}}.

\bibitem{DUNE:2020ypp}
{\bfseries DUNE} Collaboration, B.~Abi {\em et~al.}, ``{Deep Underground
  Neutrino Experiment (DUNE), Far Detector Technical Design Report, Volume II:
  DUNE Physics},'' \href{http://arxiv.org/abs/2002.03005}{{\ttfamily
  arXiv:2002.03005 [hep-ex]}}.

\bibitem{Planck:2018vyg}
{\bfseries Planck} Collaboration, N.~Aghanim {\em et~al.}, ``{Planck 2018
  results. VI. Cosmological parameters},''
  \href{http://dx.doi.org/10.1051/0004-6361/201833910}{{\em Astron. Astrophys.}
  {\bfseries 641} (2020) A6}, \href{http://arxiv.org/abs/1807.06209}{{\ttfamily
  arXiv:1807.06209 [astro-ph.CO]}}. [Erratum: Astron.Astrophys. 652, C4
  (2021)].

\bibitem{Euclid:2024imf}
{\bfseries Euclid} Collaboration, M.~Archidiacono {\em et~al.}, ``{Euclid
  preparation. Sensitivity to neutrino parameters},''
  \href{http://arxiv.org/abs/2405.06047}{{\ttfamily arXiv:2405.06047
  [astro-ph.CO]}}.

\bibitem{Project8:2022wqh}
{\bfseries Project 8} Collaboration, A.~A. Esfahani {\em et~al.}, ``{The
  Project 8 Neutrino Mass Experiment},'' in {\em {Snowmass 2021}}.
\newblock 3, 2022.
\newblock \href{http://arxiv.org/abs/2203.07349}{{\ttfamily arXiv:2203.07349
  [nucl-ex]}}.

\bibitem{KamLAND-Zen:2024eml}
{\bfseries KamLAND-Zen} Collaboration, S.~Abe {\em et~al.}, ``{Search for
  Majorana Neutrinos with the Complete KamLAND-Zen Dataset},''
  \href{http://arxiv.org/abs/2406.11438}{{\ttfamily arXiv:2406.11438
  [hep-ex]}}.

\bibitem{LEGEND:2021bnm}
{\bfseries LEGEND} Collaboration, N.~Abgrall {\em et~al.}, ``{The Large
  Enriched Germanium Experiment for Neutrinoless $\beta\beta$ Decay}:
  {LEGEND-1000 Preconceptual Design Report},''
  \href{http://arxiv.org/abs/2107.11462}{{\ttfamily arXiv:2107.11462
  [physics.ins-det]}}.

\bibitem{nEXO:2021ujk}
{\bfseries nEXO} Collaboration, G.~Adhikari {\em et~al.}, ``{nEXO: neutrinoless
  double beta decay search beyond 10$^{28}$ year half-life sensitivity},''
  \href{http://dx.doi.org/10.1088/1361-6471/ac3631}{{\em J. Phys. G} {\bfseries
  49} no.~1, (2022) 015104}, \href{http://arxiv.org/abs/2106.16243}{{\ttfamily
  arXiv:2106.16243 [nucl-ex]}}.

\bibitem{Hyper-Kamiokande:2018ofw}
{\bfseries Hyper-Kamiokande} Collaboration, K.~Abe {\em et~al.},
  ``{Hyper-Kamiokande Design Report},''
  \href{http://arxiv.org/abs/1805.04163}{{\ttfamily arXiv:1805.04163
  [physics.ins-det]}}.

\bibitem{Alekou:2022emd}
A.~Alekou {\em et~al.}, ``{The European Spallation Source neutrino super-beam
  conceptual design report},''
  \href{http://dx.doi.org/10.1140/epjs/s11734-022-00664-w}{{\em Eur. Phys. J.
  ST} {\bfseries 231} no.~21, (2022) 3779--3955},
  \href{http://arxiv.org/abs/2206.01208}{{\ttfamily arXiv:2206.01208
  [hep-ex]}}. [Erratum: Eur.Phys.J.ST 232, 15--16 (2023)].

\bibitem{Ding:2011cm}
G.-J. Ding, L.~L. Everett, and A.~J. Stuart, ``{Golden Ratio Neutrino Mixing
  and $A_5$ Flavor Symmetry},''
  \href{http://dx.doi.org/10.1016/j.nuclphysb.2011.12.004}{{\em Nucl. Phys. B}
  {\bfseries 857} (2012) 219--253},
  \href{http://arxiv.org/abs/1110.1688}{{\ttfamily arXiv:1110.1688 [hep-ph]}}.

\bibitem{Li:2015jxa}
C.-C. Li and G.-J. Ding, ``{Lepton Mixing in $A_5$ Family Symmetry and
  Generalized CP},'' \href{http://dx.doi.org/10.1007/JHEP05(2015)100}{{\em
  JHEP} {\bfseries 05} (2015) 100},
  \href{http://arxiv.org/abs/1503.03711}{{\ttfamily arXiv:1503.03711
  [hep-ph]}}.

\bibitem{Novichkov:2018nkm}
P.~P. Novichkov, J.~T. Penedo, S.~T. Petcov, and A.~V. Titov, ``{Modular
  A$_{5}$ symmetry for flavour model building},''
  \href{http://dx.doi.org/10.1007/JHEP04(2019)174}{{\em JHEP} {\bfseries 04}
  (2019) 174}, \href{http://arxiv.org/abs/1812.02158}{{\ttfamily
  arXiv:1812.02158 [hep-ph]}}.

\bibitem{Ding:2019xna}
G.-J. Ding, S.~F. King, and X.-G. Liu, ``{Neutrino mass and mixing with $A_5$
  modular symmetry},''
  \href{http://dx.doi.org/10.1103/PhysRevD.100.115005}{{\em Phys. Rev. D}
  {\bfseries 100} no.~11, (2019) 115005},
  \href{http://arxiv.org/abs/1903.12588}{{\ttfamily arXiv:1903.12588
  [hep-ph]}}.

\end{thebibliography}

\providecommand{\href}[2]{#2}\begingroup\raggedright\endgroup

\end{document}